\documentclass[fleqn,usenatbib]{mnras}

\usepackage{newtxtext,newtxmath}
\newcommand{\ascii}[1]{{\fontfamily{qcr}\selectfont #1}}
\usepackage{multirow}
\usepackage{float}

\usepackage[T1]{fontenc}

\DeclareRobustCommand{\VAN}[3]{#2}
\let\VANthebibliography\thebibliography
\def\thebibliography{\DeclareRobustCommand{\VAN}[3]{##3}\VANthebibliography}

\usepackage{graphicx}	
\usepackage{amsmath}	
\usepackage{xcolor}

\title[X-ray AGN in Boötes]{X-ray AGN in Boötes: The lack of growth of the most massive black holes\\ since $z=4$}

\author[P. Guetzoyan et al.]{Paloma Guetzoyan$^{1}$\thanks{E-mail: paloma.guetzoyan@ed.ac.uk},
James Aird$^{1}$,
Antonis Georgakakis$^{2}$,
Alison L. Coil$^{3}$,
Cassandra Barlow-Hall$^{1}$,
\newauthor
Ryan C. Hickox$^{4}$,
Amy L. Rankine$^{1}$,
and Bryan A. Terrazas$^{5}$ 
\\
$^{1}$Institute for Astronomy, University of Edinburgh, Royal Observatory, Edinburgh EH9 3HJ, UK\\
$^{2}$IAASARS, National Observatory of Athens, GR-15236 Penteli, Greece\\
$^{3}$Department of Astronomy and Astrophysics, University of California, San Diego, 9500 Gilman Drive, La Jolla, CA 92093, USA\\
$^{4}$Department of Physics and Astronomy, Dartmouth College, 6127 Wilder Laboratory, Hanover, NH 03755, USA\\
$^{5}$Department of Physics \& Astronomy, Oberlin College, Oberlin, OH, 44074, USA
}

\date{Accepted 2024 November 9. Received 2024 October 18; in original form 2024 August 19}

\pubyear{2024}

\begin{document}
\label{firstpage}
\pagerange{\pageref{firstpage}--\pageref{lastpage}}
\maketitle

\begin{abstract}
Supermassive Black Holes (BHs) are known to efficiently grow through gas accretion, but even sustained and intense mass build-up through this mechanism struggles to explain the assembly of the most massive BHs observed in the local Universe. Using the \textit{Chandra} Deep-Wide Field Survey (CDFWS) in the Boötes field, we measure BH--galaxy assembly in massive galaxies ($M_\star\gtrsim10^{10}\rm M_\odot$) through the AGN fraction and specific Black Hole accretion rate (sBHAR) distribution as a function of redshift and stellar mass. 
We determine stellar masses and star formation rates for a parent sample of optically selected galaxies as well as those with X-ray detections indicating the presence of an AGN through Spectral Energy Distribution (SED) fitting.
We derive a redshift-dependent mass completeness limit and extract X-ray information for every galaxy as to provide a comprehensive picture of the AGN population in massive galaxies. 
While X-ray AGN samples are dominated by moderately massive host galaxies of $M_{\star} \geqslant 10^{10}\rm M_{\odot}$, we do not find a strong stellar mass dependence in AGN fraction (to limits in sBHAR), indicating a bias towards massive galaxies in the observed samples. We derive BH-galaxy growth tracks over time, which reveal that while most BH mass has been accumulated since $z=4$ for lower mass BHs, the assembly of the most massive BHs is more complex, with little to no relative mass gain since $z=4$, implying that rapid and intense growth episodes prior to $z=4$ were necessary to form these massive BHs.

\end{abstract}




\section{Introduction}

Supermassive Black Holes (BHs) of mass $M_{BH} = 10^{6-10}\, \rm M_{\odot}$ are found at the centre of most massive local galaxies \citep{kormendy_inward_1995, kormendy_supermassive_2001} and are known to power Active Galactic Nuclei (AGN) through intense accretion of gas and dust. Through this accretion process and the associated outflows, AGN can release a remarkable amount of energy \citep{shapiro_black_1983}, manifested by strong radiation from radio to X-ray, sometimes outshining the host galaxy. Under their most luminous form, AGN are called Quasars (QSO) and can be detected up to very high redshift. Large QSO populations are now being revealed at high redshifts $ z\geqslant\, 7$ only a few million years after the Big Bang \citep{banados_800-million-solar-mass_2018, wang_luminous_2021}, as well as recent JWST discoveries confirming the existence of high-z AGN that appear to be powered by BHs with $M_{BH}\gtrsim10^6-10^9$ $\mathrm{M_\odot}$ \citep[e.g.][]{kocevski_hidden_2023, kokorev_uncover_2023,marshall_ga-nifs_2023,maiolino_small_2024, greene_uncover_2024}. This discovery brought to light an elusive AGN population at high redshift, and showed that massive BHs are already a common feature of galaxies within the first $\sim$Gyr of cosmic time. This raises the question of how those high redshift BHs assembled and grew so quickly. Having already assembled a significant amount of mass very early on, will these black holes continue to grow at later cosmic times, and if so, how much of the massive BH population is newly built up at later redshifts? Another remaining question is how does this late assembly of BHs proceed alongside the build up of the galaxy population?

It has been widely accepted that properties of BHs and galaxies appear to be correlated but the physical nature of this connection remains unclear \citep{kormendy_coevolution_2013}. Strong evidence shows that both BH growth and galaxy growth happens in tandem, suggesting that one would drive the build-up of the other. This co-evolution is supported, for example, by the similar evolution of the star formation rate density and AGN accretion density over cosmic time, which shows that the total amount of both BH and galaxy growth follow the same evolutionary trend through time \citep{madau_high_1996,boyle_cosmological_1998,delvecchio_tracing_2014,aird_evolution_2015}. In addition to this correlation seen across all cosmic time, local observations reveal a tight instantaneous connection between the properties of BHs and galaxies. Empirical scaling relations between BH mass and galaxy stellar mass indicate that the more massive a galaxy is, the more massive the BH at its centre \citep[][]{kormendy_coevolution_2013, greene_intermediate-mass_2020,reines_relations_2015}. 
Furthermore, BH mass seems to be tightly linked to other galaxy properties such as bulge luminosity \citep{magorrian_demography_1998, gultekin_m-sigma_2009}, velocity dispersion in the bulge \citep{ferrarese_fundamental_2000,gebhardt_black_2000,tremaine_slope_2002}, bulge mass \citep{haering_black_2004,marconi_relation_2003}, or star formation properties \citep{martin-navarro_stellar_2016, terrazas_quiescence_2016, piotrowska_quenching_2022}. We gain insight into how galaxies have assembled through these properties, as well as the assembly of their central BHs. Indeed, BH mass is the remnant of past AGN activity and is thus the observable evidence of earlier periods of growth. Thus, BH mass correlations with galaxy properties indicate a link between this mass build up and the build up of the galaxy itself. These empirical relations support the idea of a co-evolution between BHs and galaxies across all cosmic time, and the nature of this connection still needs to be investigated, especially  how and when BHs have assembled within the high mass galaxy population.

An effective way of identifying AGN is to use deep X-ray surveys, mapping the very hot and energetic components of the Universe. Though AGN emit radiation across the whole electromagnetic spectrum, the light might be absorbed at certain wavelengths by dust or gas along the line-of-sight, either from the roughly toroidal obscurer that surrounds the accretion disc \citep{urry_unified_1995}, other gas and dust along the line-of-sight associated with an AGN-driven outflow \citep{honig_redefining_2019}, or gas within the interstellar medium of the host galaxy \citep{gilli_supermassive_2022}. 
However, X-rays are much less impacted by this absorption and are ideal to reveal an obscured AGN population not seen at other wavelengths \citep{brandt_cosmic_2015, netzer_revisiting_2015,hickox_obscured_2018}. 
X-ray selection also allows us to build a census of the AGN population at all redshifts, regardless of their relative luminosity to their galaxy. Indeed, X-rays offer the best observational contrast between AGN and galaxy, and act as a probe to identify weaker AGN that would have been washed out by the host galaxy emission at other wavelengths. Additionally, X-ray emission provides a tracer of BH growth, meaning that X-ray detected galaxies are host to currently accreting BHs and provide an efficient means to trace BH-galaxy assembly.

The entanglement of BH-galaxy evolution has long been studied and one particular aspect of this relation is the link between AGN properties and host stellar mass $M_{\star}$. One still debated question is whether AGN are preferentially found in more massive galaxies. \cite{aird_primus_2012} found that the probability of a galaxy hosting an AGN of a given X-ray luminosity increases strongly with stellar mass. However, BHs in different stellar mass hosts, with correspondingly distinct BH masses and very different accretion rates, may produce the same observed luminosity. When interpreting the same results using specific accretion rate, i.e X-ray luminosity normalized by stellar mass, no evidence for a preferential mass range was found and it was shown that AGN are prevalent in galaxies of all mass $M_{\star} \geqslant \, \rm 10^{9.5}\, M_{\odot}$. Similarly, \cite{bongiorno_accreting_2012} found that the probability of a galaxy to host an AGN is also mass independent. These studies concluded that the prevalence of high-mass host galaxies in X-ray selected AGN samples is primarily a selection effect, introducing an observational bias towards high-mass galaxies at fixed accretion rates \citep[see also][]{birchall_incidence_2022}. However, more recent studies have shown different results when  moving toward higher redshift. \cite{aird_x-rays_2018} found that the probability of star-forming galaxies hosting an AGN of a given specific accretion rate is indeed mass dependent, in line with the results of \cite{georgakakis_observational_2017,yang_linking_2018} who also found a mass dependency in the specific accretion rate distributions.

Despite many studies on the subject, the AGN incidence within galaxies and more specifically as a function of stellar mass remains unclear and raises a few open questions to be addressed. Here, we will focus on the X-ray AGN population within high-mass galaxies ($M_{\star} \geqslant 10^{10}\, \rm M_{\odot}$) to place improved constraints on the specific accretion rate distribution and the AGN fraction in this regime. Galaxies of such masses and their BHs are thought to assemble their mass in tandem, with BHs growing through accretion---alternating between on-and-off AGN phases---while galaxies grow more gradually through star formation. Once averaged over sufficiently long timescales it can be demonstrated that galaxies and black holes grow with one another \citep{hickox_black_2014}; however, this dual process is not guaranteed to hold at the high-mass end of the $M_{BH} - M_{\star}$ space. A large fraction of the high-mass population has quenched in star formation by $z\leqslant 1$, yet it may be the AGN activity and thus BH growth continues \citep{aird_agn_2022, ni_incidence_2023}. Conversely, despite ongoing AGN activity, growing the most massive BHs found within massive galaxies by $z=0$ remains challenging. Recent JWST results, by finding large populations of AGN at high redshifts in hosts with very low stellar masses relative to their BH \citep[e.g.][]{larson_ceers_2023}, suggest that the bulk of BH mass build up within these galaxies may occur very early, before the stellar mass is assembled.
This paper aims to directly track the later assembly of BH mass within the massive galaxy population.

We will rely on the \textit{Chandra} Deep-Wide Field Survey (CDWFS) in the Boötes field \citep{masini_chandra_2020}, probing a wide and deep area of 9.3 $\rm deg^2$ with 6891 AGN identifications up to very high redshift thanks to the excellent sensitivity of \textit{Chandra} and a large total exposure time of 3.4 $\rm Ms$. This survey offers an impressive combination of deep imaging to unveil faint sources while also spanning a wide area needed to provide the necessary statistics within higher mass galaxies over a broad range in redshift.
The X-ray detections have been matched to the multi-wavelength catalogue in Boötes \citep{kondapally_lofar_2021} to provide host galaxy properties and place these AGN within the context of the galaxy population.

In this paper, we investigate the AGN incidence in massive galaxies across a broad range of redshifts to look at how BH growth behaves in this high mass population. In particular, we aim to quantify how much BH assembly is still on-going for already very massive BHs and how it relates to the mass of the host galaxy. In Section \ref{sec:data}, we briefly describe the datasets used both in the optical/IR to construct a sample of galaxies and in the X-ray to identify their AGN. In Section \ref{sec:sample}, we describe our methods to clean our sample from stars, then proceed to the derivation of stellar masses through SED fitting while accounting for any contamination by the AGN light. With these estimates, we then compute a redshift-dependent mass completeness limit to exclude under-represented low-mass galaxies from our sample. Finally, in Section \ref{sec:results}, we present our various results showing stellar mass evolution with redshift for both optically selected galaxies and X-ray selected AGN from the optical parent sample, X-ray luminosity potential dependence on stellar mass as well as AGN fractions as a function of stellar mass and redshift after careful corrections for the X-ray completeness. We end by showing final derivations of the specific accretion rate distributions in stellar mass and redshift bins. We discuss our findings and summarize our results in Section \ref{sec:discussion} and \ref{sec:conclusion} respectively.

All results are derived under the standard cosmology assumptions ($H_0 = 70 \,\rm km/s/Mpc, \Omega_m = 0.3, \Omega_\lambda = 0.7$).

\section{Data}
\label{sec:data}

In this section, we describe the data used in our multi-wavelength analysis of the Boötes field. We benefit from a large wavelength coverage in the optical to IR (section \ref{sec:data_optIR}) as well as dedicated \textit{Chandra} observations (section \ref{sec:data_X}). 

\subsection{Optical/IR photometry}
\label{sec:data_optIR}
We focus our study on the Boötes field located at $\rm (\alpha,\delta) = (14h32m00s, +34^{\circ}30'00")$, covering a substantial area of approximately 9.3~deg$^2$. Extensive photometry was obtained in this region by combining NOAO Deep Wide Field Survey 
\citep[NDWFS:][]{jannuzi_noao_1999}
for deep optical data with the $B_w, R, I$ and $K$ bands, GALEX surveys in the $NUV$ and $FUV$ bands (from data release 6/7\footnote{\url{https://galex.stsci.edu/GR6/}}), $z$-band from \cite{cool_zbootes_2007}, near-IR $J, H$ and $K_s$ bands from NEWFIRM \citep{gonzalez_2010}, and mid-IR data from the Spitzer Deep Wide Field Survey \citep[SDWFS;][]{ashby_spitzer_2009} including the Infrared Array Camera (IRAC) channels 1, 2, 3 and 4 bands (hereafter, $ch1, ch2, ch3$ and $ch4$). The full details of the creation of the multi-wavelength catalogue are given by \cite{kondapally_lofar_2021}.

For redshift information, we relied on spectroscopic measurements from the AGN and Galaxy Evolution Survey \citep[AGES;][]{kochanek_ages_2012}. In cases where spectroscopic redshifts were not available or of insufficient quality ($\rm Q <\, 3$), we supplemented our dataset with photometric redshift estimates from \cite{duncan_lofar_2021}. Comparing to the available spec-$z$, they found an outlier fraction of $\approx 1.5 - 1.8 \%$ for galaxies and $18 - 22 \%$ for AGN, as well as a robust scatter, i.e the median deviation, of $1.6 - 2 \%$ and $6.4 - 7 \%$ for respective subsets. Our sample spans a redshift range of $z \leqslant \, 7$, with a limiting $R$ magnitude of $R \leqslant \, 25$.

\subsection{X-ray imaging}
\label{sec:data_X}
The Boötes region has not only been extensively studied in optical wavelengths but also in the deep X-ray domain, thanks to the \textit{Chandra} Deep-Wide Field Survey \citep[CDWFS,][]{masini_chandra_2020}. Over the past decade, significant efforts have been made to update and enhance the parameter space probed by different \textit{Chandra} surveys, either by improving the flux limitations or expanding the survey areas. The CDWFS in Boötes was specifically designed to address a sensitivity gap in terms of moderate depth coverage over multiple deg$^2$ regions of sky. To achieve this goal, a combination of previous \textit{Chandra} observations of depth $5\, \rm ks$ \citep{murray_xbootes_2005} were supplemented by a recent (cycle 18) program focusing on increasing the depths of the central $\sim$6~deg$^2$ of the Boötes field.
The total observing time for this program amounted to 1 Ms, with an average exposure time of around 30 ks per pointing. The survey encompasses a total of 281 pointings, acquired between 2003 and 2018.

The acquired X-ray data underwent rigorous reduction and analysis using the \textit{Chandra} Interactive Analysis of Observations (CIAO) software version 4.11 \citep{fruscione_ciao_2006}, coupled with the Calibration Database (CALDB) version 4.8.2. A comprehensive description of the data reduction process is given by \cite{masini_chandra_2020}. X-ray sources are identified in three energy bands: full $\rm (0.5 - 7.0\, keV)$, soft $\rm (0.5 - 2.0\, keV)$, and hard $\rm (2.0 - 7.0\, keV)$ with limiting fluxes of $4.7 \times 10^{-16}, 1.5 \times 10^{-16}$, and $9 \times 10^{-16} \, \rm erg/s/cm^2$ respectively. After removing spurious sources, i.e. sources returned by a simulated survey which don't correspond to a real detection due to background fluctuations, the survey identified in total 6891 X-ray sources. Spurious sources are removed based on a probability threshold $P$, where the probability of a source being a real detection is higher below this threshold ($\log P = -4.63, -4.57, -4.40$, in the full, soft and hard bands). In this paper, we only consider sources detected in the hard band. We also include sources without significant X-ray detections by extracting X-ray data at each position from the optical catalogue. At each position, we retrieve total counts $N$, background counts $B$, effective exposures, the size of the Point Spread Function (PSF), and the varying Energy Conversion Factor (ECF) that accounts for the deterioration of \textit{Chandra}'s soft energy response over the 15 years between the first and last pointings. We then convert the net source counts ($S=N-B$) into counts per exposure time, i.e count rates, that are converted into X-ray fluxes based on the ECF. We also correct for the 90 $\%$ enclosed energy fraction of \textit{Chandra} PSF. This extraction process allows us to use both the X-ray detected sources as well as the X-ray information of non-detected sources to push down the limiting sensitivity of the survey. 

In X-ray photometry, particular care must be taken to correct for the Eddington bias. The Eddington bias \citep{eddington_correction_1940} stems from the non-uniform scattering of detections from their true fluxes. 
Faint sources that appear brighter due to positive fluctuations from photometric noise outnumber bright sources with under-estimated fluxes, introducing a bias where sources close to the flux limit tend to have overestimated fluxes. To correct for this bias, we follow the method described in \cite{laird_aegis-x_2009}, where they approximate the X-ray source number density distribution as a power-law with a  slope $\beta = 1.5$ at faint fluxes and use it as a prior on the Poisson likelihood that describes the observed X-ray photon counts. The best posterior estimate of the net source counts, $S$, corresponding to the mode of the posterior distribution is then given by

\begin{eqnarray}
    S = \frac{1}{2} \left(N-B+\beta + \sqrt{(N-B+\beta)^2+4B\beta}\right)
\end{eqnarray}
which is subsequently converted to a count rate (based on the effective exposure) and an updated X-ray flux estimate (based on the ECF and 90\% enclosed energy fraction) as described above.

Moreover, one must account for survey completeness and correct for the observational bias towards high-flux sources that have a higher detection probability than low-fluxes sources due to sensitivity limitations. \cite{masini_chandra_2020} derived the \textit{Chandra} sensitivity curves in each band by simulating the observed data in the Boötes field and comparing the output and input fluxes to derive the completeness of the survey as a function of flux. If the simulation recovers all sources at a given flux, the survey is complete and gives a detection probability of 1. Rescaled by the total area of the survey, this creates the sensitivity curve used to correct for survey completeness (see Section \ref{sec:acurve} for description of the correction method).

To derive intrinsic rest-frame luminosities in the 2-10 keV band, we employ the following procedure. We first retrieve the observed fluxes, $f_{obs}$, in the \textit{Chandra} hard band (2 - 7 keV). We then convert these observed fluxes into intrinsic fluxes, $f_{int}$, using the correction factor k computed by \cite{masini_chandra_2020}, which is given by $f_{int} = f_{obs}/k$ and corrects for Galactic absorption due to the hydrogen column density along the line of sight $N_H$. The next step, involves converting the intrinsic fluxes to rest-frame by adopting a typical power-law with a spectral index of $\Gamma = 1.9$. The rest-frame intrinsic flux is computed as $f_{int,RF} = f_{int}(1+z)^{\Gamma - 2}$, where $z$ represents the spectroscopic redshift from the AGES survey or the photometric redshift estimates from \cite{duncan_lofar_2021} if the former is not available in good quality. Having obtained the intrinsic rest-frame flux, we then proceed to calculate the intrinsic rest-frame luminosity, denoted as $L_{int,RF}$. The luminosity is computed from the flux such that $L_{int,RF} = 4\pi D_L^2f_{int,RF}$, where $D_L$ represents the luminosity distance derived under the standard cosmology assumptions ($H_0 = 70$\,km\,s$^{-1}$\,Mpc$^{-1}$, $\Omega_m = 0.3$).

Finally, we convert the intrinsic rest-frame luminosity to the 2-10 keV band:

\begin{eqnarray}
    L_{(2-10)\rm keV} = L_{int,RF} \frac{10^{2-\Gamma}- 2^{2-\Gamma}}{7^{2-\Gamma}- 2^{2-\Gamma}}
\end{eqnarray}

The X-ray and optical samples were already cross-matched by \citet{masini_chandra_2020}, so both catalogues were science ready for data analysis. The cross-matching was done using NWAY \citep{salvato_finding_2018} which uses a Bayesian approach to match the X-ray and I-band catalogues. X-ray sources and optical counterparts are associated based on separation distance and magnitude. The majority of the counterparts were found within a radius of $\rm 2"$. $\rm 85\%$ of matches (5852 sources out of 6891) are robust optical associations. In the following, we examine two populations: the optically selected galaxies in Boötes and the subset of these galaxies that are identified as containing X-ray selected AGN.

\section{Building a mass-complete sample}
\label{sec:sample}
In this section, we describe the procedure adopted to build a reliable and stellar-mass-complete galaxy sample from our photometric sample. We start by removing stellar objects using a colour-colour selection described in Section \ref{sec:stellar_cut}, then proceed to derive stellar masses for both our parent sample and X-ray detected galaxies through SED fitting as detailed in Section \ref{sec:SED}. 
In Section \ref{sec:mass_comp} we define a redshift-dependent mass completeness limit to discard low-mass galaxies under-represented in our sample due to the limits of the optical imaging. 

\subsection{Cleaning and Removing stars}
\label{sec:stellar_cut}

\begin{figure}
    \centering
    \includegraphics[width=\columnwidth]{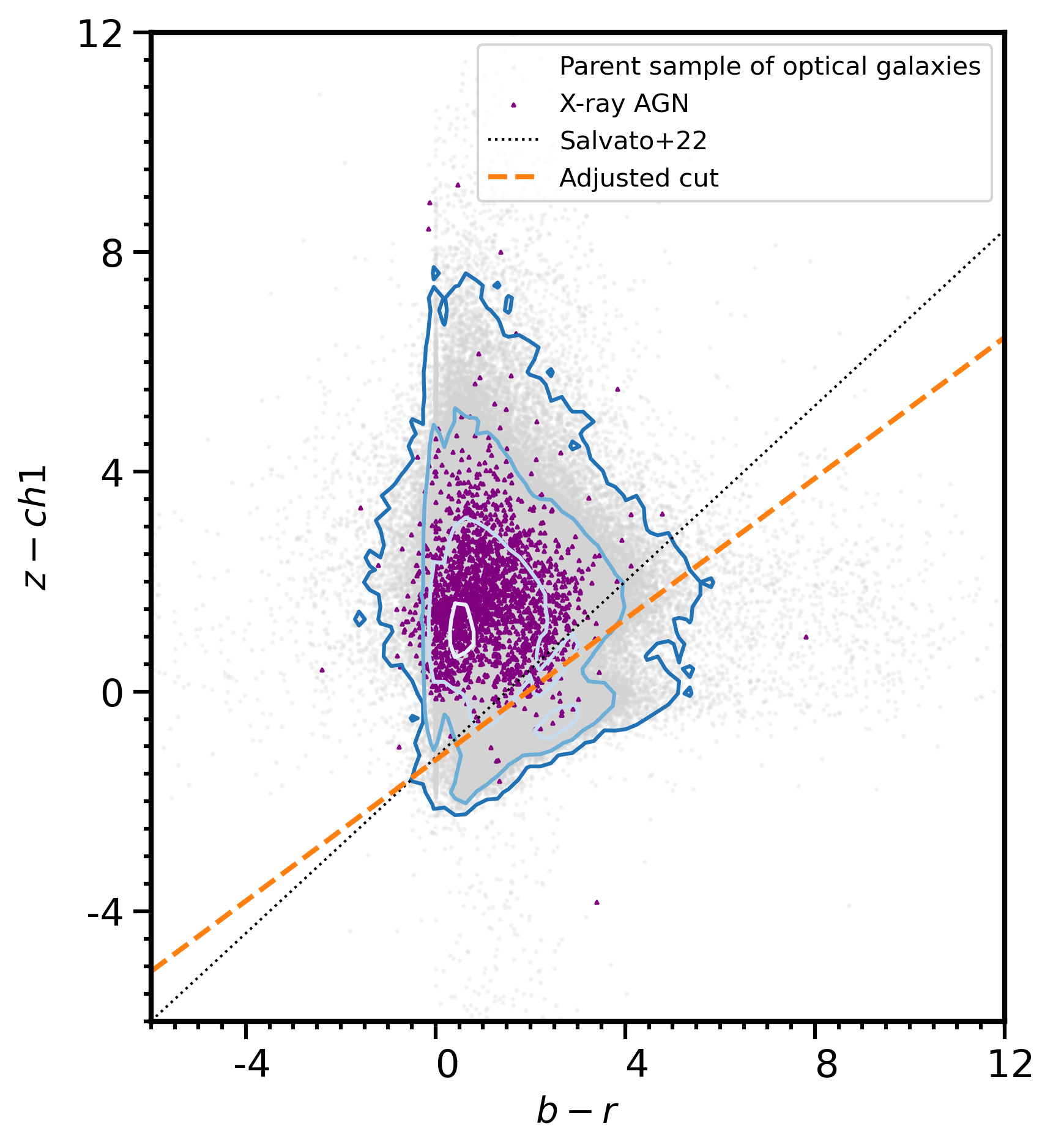}
    \caption{Colour-colour diagram identifying the stellar locus below the orange dashed line for the optical sample and X-ray detected sources in grey and purple. Our cut shown in orange is adjusted from the \citet{salvato_erosita_2022} fit in black to match the correct magnitudes. Our adjusted relation is thus effective at separating stars (that lie along a clear locus) from the larger galaxy/AGN populations.
    }
    \label{fig:star_cut}
\end{figure}

The first step in cleaning our sample is to apply the recommended flag cuts as advised in \cite{kondapally_lofar_2021}. We keep sources with complete multi-wavelength coverage that are not contaminated by bright stars (i.e. requiring OVERLAP=1 and CLEAN=1 flags). We also make a cut at $R <\rm 25$ corresponding to the depth of our survey, so as not to include spuriously detected faint objects.

Removing stars can be done based on photometry, as described in \cite{salvato_erosita_2022}, where they derived a cut based on $(z-w1)$ and $(g-r)$ colours. We adjusted this cut to use bands available for our photometric coverage, and derived a new parametrization of $(b-r)$ with $(z-ch1)$. We assume a power-law spectral energy distribution of varying index $\beta$, that we use to convert between $(g-r)$ colour \citep[as used by][]{salvato_erosita_2022} and the corresponding $(b-r)$ given the different central wavelengths of the band passes ($\lambda_b = 418\, \rm nm, \lambda_g = 464\, \rm nm$). We do the same to convert $(z-w1)$ (where $w1$ indicates the WISE channel 1 band pass of central wavelength $\lambda_{w1} = 3352.6\, \rm nm$) to a $(z-ch1)$ colour (where $ch1$ corresponds to the IRAC channel 1 band pass, $\lambda_{ch1} = 3537.8\, \rm nm$). We implement these new expressions in the calibration derived in \cite{salvato_erosita_2022} and we find that stars can be separated from galaxies based on,

\begin{eqnarray}
    (z-ch1) &<& 0.64\times (b-r) - 1.25.
    \label{eq:new_starcut}
\end{eqnarray}

In Figure \ref{fig:star_cut}, we show the distribution of optical sources and the X-ray detected subset in our revised colour-colour plot. The black dotted line is the stellar delimitation as in \cite{salvato_erosita_2022} while the orange dashed line shows Equation \ref{eq:new_starcut}. Sources below the line are classified as stars and are thus removed from our sample. For both axes, a greater colour value indicates a redder source, in the infrared for $(z-ch1)$ and in the optical for $(b-r)$. This photometric classification shows that galaxies tend to have redder $(z-ch1)$ colours, while at fixed $(z-ch1)$ colour, stars tend to be redder.
We note that relatively few X-ray sources lie below the cut, in the stellar region; the majority of X-ray point sources are AGN. 
We assume that sources lacking detections in either $z$ or $ch1$ (and thus cannot be placed on this diagram), which in general are optically fainter, are galaxies rather than stars and thus retain them in our sample. We performed a careful cross-matching with the Gaia DR3 sources \citep{gaia_collaboration_gaia_2023}, before and after our stellar cut and found that we remove over $98\%$ of Gaia stars in our original sample using our colour selection. The $0.2\%$ sources remaining have extremely low parallaxes and proper motions centered on 0.

\subsection{SED fitting}
\label{sec:SED}
In order to derive physical properties, such as stellar masses, we employed a process called spectral energy distribution (SED) fitting, making use of the extensive photometric data acquired in the Boötes region. 
This comprehensive wavelength coverage survey allows us to measure the SEDs of galaxies across a wide range of redshifts and constrain the contributions from different processes including stars, dust, and active galactic nuclei (AGN).

Many codes are publicly available to perform SED fitting. Here, we chose to use the \textsc{python} Code Investigating GALaxy Emission \citep[\textsc{cigale};][]{boquien_cigale_2019}, making use of its short computing time and low storage needs as well as innovative additional AGN component in the model of SEDs. In the following, we describe the models we built and the components used to accurately fit the SEDs of our host/AGN-dominated galaxies.

We adopt the same model to describe the galaxy contribution to the SED for both the optical parent sample and the X-ray detected sub-sample, while adding an AGN component for the X-ray sample to account for the AGN contribution at optical/IR wavelengths and separate it from the stellar/dust emission of the host. In order to incorporate the temporal evolution of star formation rates (SFRs), we employed a delayed Star Formation History (SFH) model. This model offers smoother variations in SFR over time rather than the sudden onset of star formation that is assumed when modelling galaxies with exponentially declining SFH models. It is well-known that the assumption of exponentially declining SFH can introduce biases in the determination of stellar masses and SFR \citep{pacifici_evolution_2016,carnall_how_2019,leja_how_2019}, whereas a delayed model provides an almost linear increase of the SFR from the age of the onset of star formation rather than an abrupt one, producing more reliable measurements of stellar masses and SFR. It reaches a peak at $\tau_{main}$, then gradually decreases.

To describe the stellar population, we adopted the popular library defined in \cite{bruzual_stellar_2003} using a grid of fixed metallicity along with the Salpeter Initial Mass Function (IMF). To differentiate between young and old stars, we allow for an age-dependent reddening factor, so as to account for young stellar populations still embedded in their dust clouds, which absorb at short wavelengths in addition to absorption by dust in the Interstellar Medium (ISM). To account for this age dependency, we assume an attenuation law from \cite{charlot_simple_2000} which allows us to compute two different attenuation contributions: dust clouds surrounding young stars and the ISM for stars of all ages. Additionally, we accounted for nebular emission using the model developed by \cite{inoue_updated_2014}. This model considers the emission originating from the most massive stars, which ionize the surrounding gas, resulting in the re-emission of light through a series of emission lines. Dust emission is modelled as in \cite{draine_andromedas_2013}, and is separated into two components. The first component is the diffuse dust emission, heated by the global stellar population, while the second component models dust emission specifically linked to star-forming regions, where contrary to the first case, dust is heated by a variable radiation field ranging from $U_{min}$ to $U_{max} = \rm 10^7$. 

Finally, for our X-ray sample, we add an AGN component to disentangle the nuclear emission from star formation, both contributing strongly in the UV. AGN emission is modelled using three radiative components: the primary source located within the torus, the scattered emission by dust, and thermal dust emission. A set of several parameters describes this component, including the radius $r$ of the torus and the fraction of light coming from the AGN \citep{stalevski_dust_2016}.

Full details of the parameter grid used can be found in Table~\ref{cigale_param}. 
We discard any sources with unreliable SED fits, with reduced $\chi^2\ge 10$, which removes $\sim3$\% of the optical parent sample and $\sim6$\% of the X-ray selected sample. 

\subsection{Mass completeness limit}
\label{sec:mass_comp}

To minimize observational biases in our sample, we derive a redshift-dependent mass completeness limit. Since our survey targets objects down to a limiting magnitude of $R \sim 25$, we are unable to detect low mass galaxies at high redshift unless they show extremely bright UV luminosities due to intense star formation. To ensure our measurements are performed on a complete and unbiased sample, we decide to consider galaxies above a mass limit, where we are confident we can detect all of them independently of their star formation rates.

To derive these mass limits, in each of our redshift bins (as adopted in Section~\ref{sec_fagn_LX} below), we take the most massive 90\% of galaxies in the $R$-band limited optical sample (after removing stars as described in Section~\ref{sec:stellar_cut} above) and derive their mass-to-light ratios, $M_\star/L$, where $L$ is the $R$-band flux of each galaxy in a redshift bin converted to luminosity using the redshift and luminosity distance under standard cosmology. 
Next, we take the 90-th greatest of these mass-to-light ratios and calculate the mass of a galaxy with this $M_\star/L$ corresponding to our limiting $R$ magnitude, $R=25$, and a redshift corresponding to the upper limit of the redshift bin. This provides a limiting mass, $M_{lim}(z)$, corresponding to the minimum mass for a galaxy that is guaranteed to enter our sample even if it has the most extreme mass-to-light ratio observed throughout the optical sample.
This process results in a smoothly evolving mass completeness limit as a function of redshift as shown in Figure \ref{fig:M-z}.

Applying this cut restricts our sample to include only more massive galaxies so that we are not biased towards the high-SFR galaxy population. We can confidently run our analysis on a complete sample at a given stellar mass for all redshifts. It is worth noting that our X-ray selected AGN are less affected by these mass completeness corrections, as they typically do not belong to the elusive category of low-stellar-mass galaxies. 

After removing stars, unreliable measurements ($\chi^2_r \geqslant\, \rm 10$) and applying our mass completeness limit, our final sample goes from 2,214,365 to 410,497 galaxies in the parent sample and from 6891 to 2114 hard X-ray selected sources with 903 AGN having spectroscopic redshift, i.e $\sim 43\%$ of the final AGN sample. The mass completeness cut ends up being the most significant cut in our optical sample, while the X-ray population is less affected by this correction (see Table \ref{tab:ncount}).

\begin{table}
    \centering
    \begin{tabular}{c|c|c}
    \multicolumn{3}{|c|}{Number count}\\
    \hline
    & Optical parent sample & X-ray selected\\
    \hline
    Total & 2214365 & 6891 \\
    \hline
    Flag cut & 1877410 & 5446 \\
    \hline
    Hard X-ray selection & - & 2716 \\
    \hline
    R cut & 1277226 & 2607 \\
    \hline
    Star cut & 1132087 & 2570 \\
    \hline
    $\chi^2_r$ cut & 1117527 & 2478 \\
    \hline
    Mass cut & 410497 & 2114 \\
    \hline
    
    \end{tabular}
    \caption{Number count of sources for the parent sample and X-ray selected sources after each cleaning step. We restrict the data to robust and reliable measurement where we are confident our sample is complete in stellar mass.}
    \label{tab:ncount}
\end{table}

\section{Results}
\label{sec:results}
In this section, we present a number of results examining the link between the stellar mass of galaxies and their X-ray emission. As all sources in our X-ray sample have $L_{(2-10)\rm keV}>10^{42}$~erg~s$^{-1}$ (see Section \ref{sec:LX}), we assume that all of these sources are AGN and that the X-ray emission is tracing the accretion rate \citep{alexander_what_2012}. 

In Section~\ref{sec:stellarmass}, we compare the average stellar masses of the X-ray AGN sample with the parent galaxy sample, whereas in
Section~\ref{sec:LX} we study the relationship (and any correlation) between the X-ray luminosity and the stellar masses for the X-ray AGN sample. 
In Section~\ref{sec:agnfraction} we measure the AGN fraction to different limits in X-ray luminosity (Section~\ref{sec_fagn_LX}) and sBHAR (Section~\ref{sec_fagn_sBHAR}) to quantify the incidence of AGN within the galaxy population and how it varies with both stellar mass and redshift. Finally, in Section \ref{sec:extract}, we present the intrinsic sBHAR probability density functions varying with stellar mass and redshift to identify the phases of BH growth through cosmic time and galaxy evolution.

\subsection{Average stellar mass of X-ray selected AGN as a function of redshift compared to the parent galaxy population}
\label{sec:stellarmass}

\begin{figure}
    \centering
    \includegraphics[width=\columnwidth]{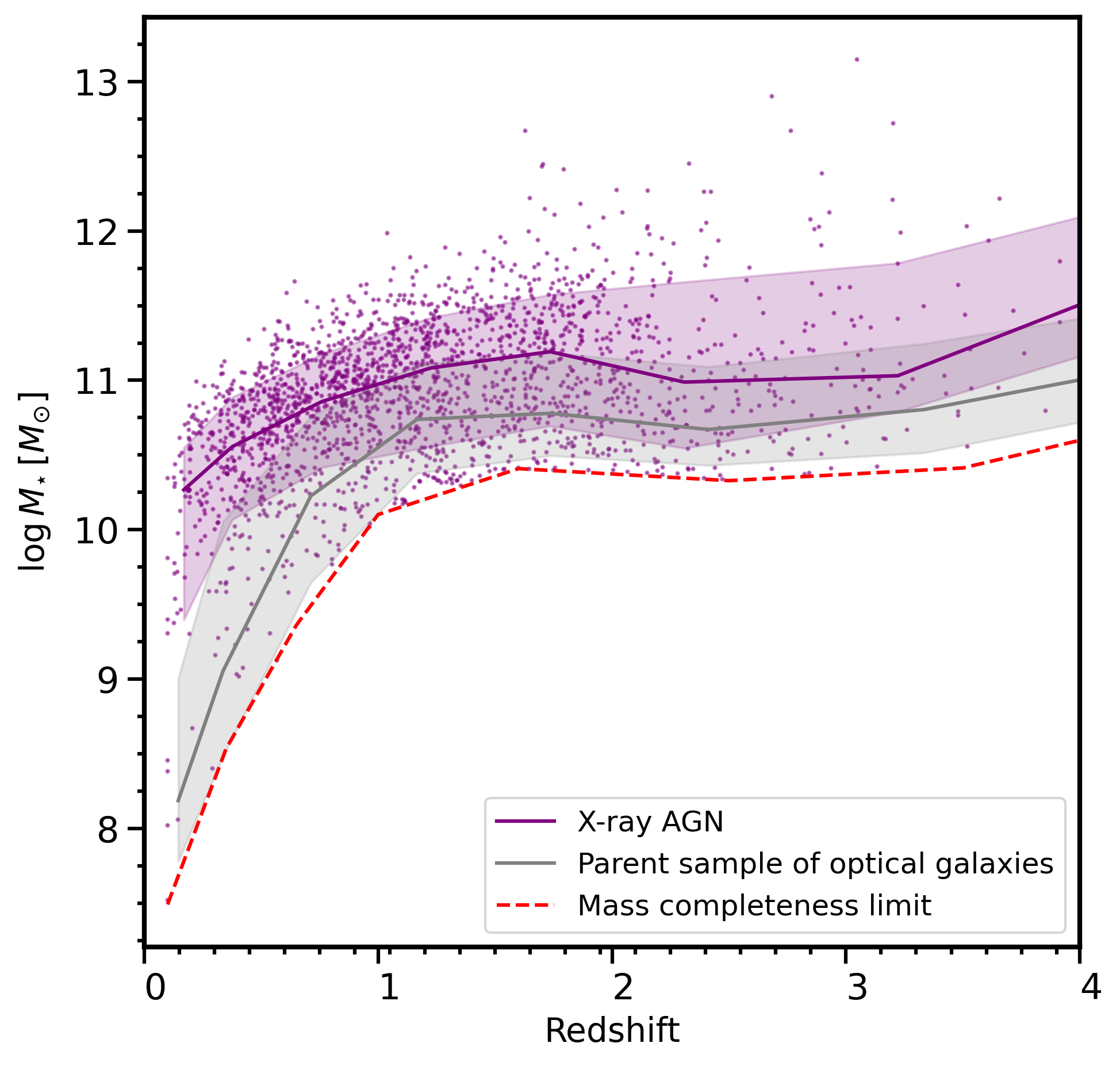}
    \caption{Distribution of the stellar masses of the optically selected parent sample of galaxies (grey) and the X-ray selected AGN identified within this sample (purple) as a function of redshift. Solid lines indicate the median stellar mass in bins of redshift while shaded areas show the $15^{th}-85^{th}$ percentile. 
    Purple points are the individual X-ray detections while individual galaxies in the optical control sample are not shown.
    The red dotted line is the redshift dependent mass completeness limit derived in Section \ref{sec:mass_comp}.
    While X-ray AGN are found across the full mass range of the parent galaxy sample, there is a strong selection bias that results in the majority of sources in a flux-limited sample being identified within the most massive galaxies at a given redshift (see Section~\ref{sec:stellarmass} for further discussion).
    }
    \label{fig:M-z}
\end{figure}

\begin{figure*}
    \centering
    \includegraphics[scale=0.85]{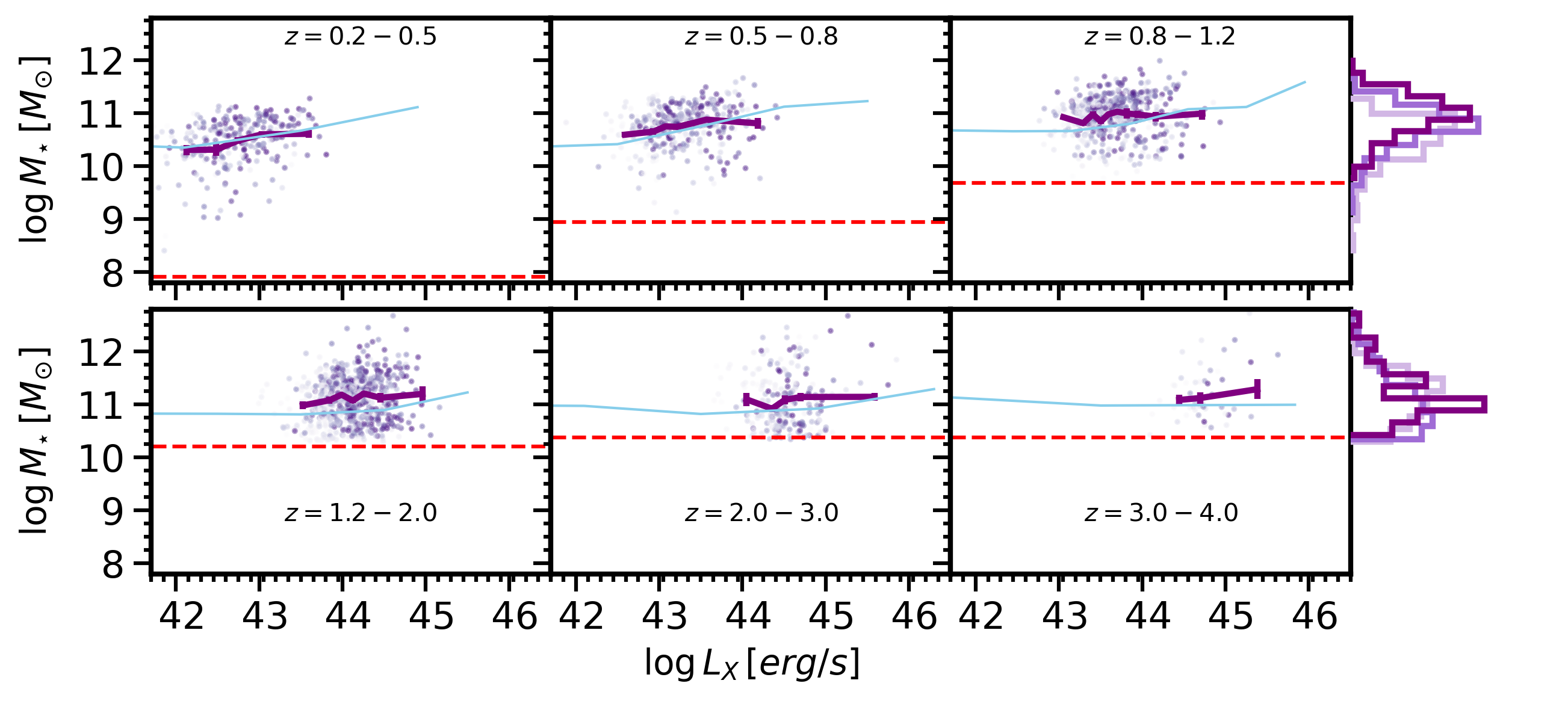}
    \caption{Total stellar mass - X-ray luminosity relation in bins of redshift for the X-ray selected AGN sample (purple dots, darker color with increasing redshift). Solid purple lines show the median $M_{\star} - L_X$ in bins of $L_X$ that we compare to the model of \citet{georgakakis_forward_2020} shown by the blue solid lines. Per Figure \ref{fig:M-z}, the red dotted lines are the mass completeness limit at the low end of each redshift bin. The change in the distributions of host stellar masses with redshift along each row is shown by the side normalized histograms, with darker shades indicating increasing redshift. 
    X-ray selected AGN are typically found to have host galaxies with masses $\sim10^{10.5}-10^{11}\, \rm M_\odot$, albeit with a very broad distribution. There is little (if any) dependence on $L_X$, although the highest luminosity AGN may be somewhat restricted to more massive host galaxies resulting in a slight upturn in the median stellar masses. 
    }
    \label{Mstar-LX}
\end{figure*}

In Figure \ref{fig:M-z}, we compare our stellar mass measurements obtained through SED fitting for the optically-selected galaxy sample and the X-ray AGN sample. While there are galaxies in our sample with redshifts out to $z\sim6$, due to the limited wavelength coverage of the optical/IR photometry we are only probing rest-frame UV wavelengths for high-redshift sources, so our stellar mass estimates start to become unreliable above $z = \rm4$ and we limit our analysis to sources below this redshift. The solid purple and grey lines show the median stellar mass in bins of redshift for our X-ray and optically-selected sources respectively while the shaded regions are the $\rm 15^{th} - 85^{th}$ percentile and thus indicate the distribution of stellar masses at a given redshift. The dashed red line is the mass completeness limit $M_{lim}(z)$ derived in Section \ref{sec:mass_comp}.

Overall, our measurements show that on average, X-ray selected AGN are preferentially found in the most massive galaxies within the parent sample of optically-selected galaxies, at all redshifts. This suggests that massive galaxies are more likely to host AGN and that X-ray selection provides an effective method for studying the co-evolution of supermassive black holes and galaxies, particularly at the high-mass end of the galaxy population. This finding could imply that rapid BH growth, characterized by strong X-ray emission, occurs predominantly in the most massive galaxies of mass $M_{\star} \geqslant 10^{10} \, \rm M_{\odot}$. However, it's important to consider that detecting AGN in the most massive galaxies could be influenced by X-ray selection biases. More massive galaxies likely host more massive BHs, which when accreting will produce a higher X-ray luminosity and as such, will dominate in flux-limited X-ray samples. Looking at the distribution of X-ray sources in the stellar mass - redshift space, the bulk of X-ray detections are indeed located in higher mass galaxies but we still find a significant number of AGN in galaxies of $M_{\star} \leqslant 10^{10}\, \rm M_{\odot}$. This distribution towards massive galaxies is mostly the result of a selection bias caused by the fact that X-ray AGN of a given accretion appear brighter in more massive galaxies, making them easier to detect.

\subsection{Dependence of host stellar mass on X-ray luminosity}
\label{sec:LX}

We investigate here the potential correlation between the host stellar masses of X-ray selected AGN and their X-ray luminosities.

In Figure \ref{Mstar-LX}, we show our measurements of stellar mass for the X-ray AGN sample as a function of X-ray luminosity for different redshift cuts, as in \cite{azadi_primus_2015}. The purple curve shows the median stellar mass in bins of X-ray luminosity. Each $L_X$ bin contains at least 50 sources for most of the redshift cuts and 100 (15), sources for the $z = \rm 1.2 - 2$, ($z =3-4$), panel. The error bars show the uncertainty in each bin computed using the bootstrapping method which is very useful to derive inferences of a limited size population. As per Figure \ref{fig:M-z}, the red dotted line shows the mass completeness limit at the low end of the redshift bin. The blue solid curve shows the model described in \cite{georgakakis_forward_2020}. This model is created by seeding galaxies drawn from the stellar mass function with AGN based on a mass-independent distribution of specific accretion rates that produces agreement with the overall X-ray luminosity function of AGN. 
This produces a mock catalogue of AGN and galaxies whose multi-wavelength properties are determined by implementing templates of SEDs. To match our sample, we apply to their mock catalogue our magnitude cut and mass completeness limit.

Looking at the median relation, we find no particular dependence of stellar mass on X-ray luminosity, for any redshift bin. 
This unvarying relation implies that similar galaxies in terms of mass can produce a very broad range of X-ray luminosities. 
Equivalently, at a fixed luminosity, we observe a broad distribution of stellar mass, as shown by the purple dots.
However, a weak upturn in the median stellar masses at high luminosity can be seen in nearly all panels, as also predicted by the \citet{georgakakis_forward_2020} model, indicating that the most luminous AGN may be restricted to only the most massive galaxies. 

 For all the redshift bins, the stellar mass distributions all peak above $\rm 10^{10}\, M_{\odot}$.
  We note that \citet{aird_primus_2013} suggest that the typical $\sim 10^{10.5}\, \rm M_{\odot}$ host masses of moderate luminosity X-ray selected AGN ($L_\mathrm{X}\sim10^{42}-10^{44}$~erg~s$^{-1}$) are a result of the relatively steep faint-end slope of the specific accretion rate distribution, combined with the characteristic break in the galaxy stellar mass function, that leads to such host masses dominating in X-ray selected samples rather than an \emph{intrinsic} preference for AGN activity to occur in more massive galaxies. 
 Our median stellar masses also
 appear to shift toward higher masses with increasing redshift (side histograms of darker colour with increasing redshift bin). However, this shift is only observed between our lowest redshift bins (top panels).
 The mass distribution at redshift $z \geqslant 1.2$ should be taken with caution as the mass completeness limit may impact the observed stellar mass distribution.

Overall, our results follow a similar trend to the model presented in \cite{georgakakis_forward_2020}, finding a mostly flat dependence of the median host stellar mass on X-ray luminosity with a very slight upturn at high $L_X$ indicating that the highest luminosity AGN are restricted to the highest mass galaxies.  We note that this upturn almost completely vanishes at high redshift when removing our mass completeness limit; some high luminosity X-ray AGN may be hosted by lower mass galaxies below our mass limit and thus the requirement of a high mass host may not be as strong as first indicated. 
The lack of a correlation between the host stellar mass and X-ray luminosity shows that AGN over a wide range of luminosities may occur in galaxies right across the mass range, induced by the broad distribution of accretion rates likely reflecting the variability of black hole growth happening on much shorter timescales than galaxy growth mechanisms.

Finally, we note that our median stellar masses are marginally higher than predicted by the \citet{georgakakis_forward_2020} model, especially at higher redshifts ($z>1.2$). This discrepancy may reflect an enhancement in the prevalence of X-ray AGN (at all luminosity levels) within the most massive galaxies, compared to the mass-independent AGN incidence in terms of specific accretion rate assumed by \citet{georgakakis_forward_2020}. 
In the following sections, we address this issue directly by measuring the incidence and  distribution of X-ray AGN in terms of their specific accretion rates within galaxies of different stellar masses. 

\subsection{Incidence and distribution of BH accretion in galaxies as a function of stellar mass and redshift}
\label{sec:agnfraction}

In this section, we start from our samples of galaxies at different stellar masses and redshifts and quantify the incidence of X-ray AGN \emph{within} such samples. 
We trace the distribution of AGN using two different tracers of accretion : the X-ray luminosity $L_X$ and the specific accretion rate $\lambda= L_X/M_{\star}$ i.e. the X-ray luminosity scaled relative to galaxy stellar mass. We compute the AGN fraction $f_{AGN}$ defined as the ratio between the number of X-ray detections (above a chosen $L_X$ or $\lambda$ limit) and the total count of galaxies within specific $M_{\star} - z$ bins to determine the prevalence of AGN within galaxies. Later on, we show our measurements of the specific BH accretion rate distributions (sBHAR) in the same $M_{\star} - z$ space. 

AGN fraction and sBHAR probability distributions are substantially influenced by the sensitivity limitations intrinsic to our X-ray survey, as well as the constraints imposed by the limiting magnitude of the parent galaxy sample. To ensure the robustness of our findings and to mitigate the impact of these limitations, we correct our measurements for X-ray completeness, following two distinct methods. We first correct for survey completeness using the \textit{Chandra} area curve in Boötes in section \ref{sec:acurve}. In section \ref{sec_fagn_LX}, we use X-ray luminosity as a tracer of BH growth to compute the AGN fraction in different luminosity cuts, and switch to specific BH accretion rate in Sections \ref{sec:extract} and \ref{sec_fagn_sBHAR} to compute the sBHAR probability distribution and AGN fraction respectively, to take into account the entanglement between stellar mass and X-ray luminosity.

\subsubsection{X-ray completeness correction}
\label{sec:acurve}

\begin{figure}
    \centering
    
    \includegraphics[width=\columnwidth]{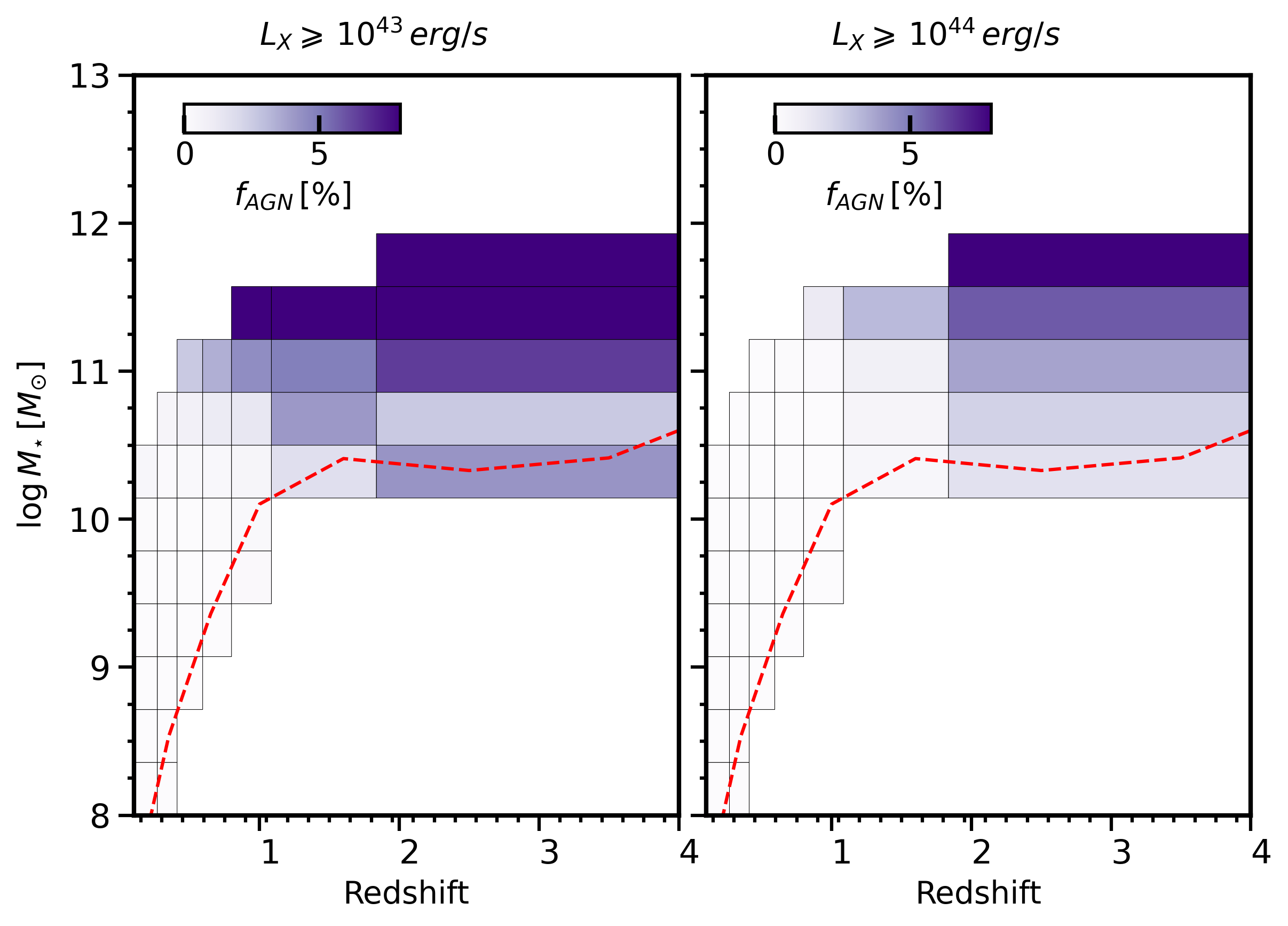}
    \includegraphics[width=\columnwidth,trim={0.3cm 0 0 0}]{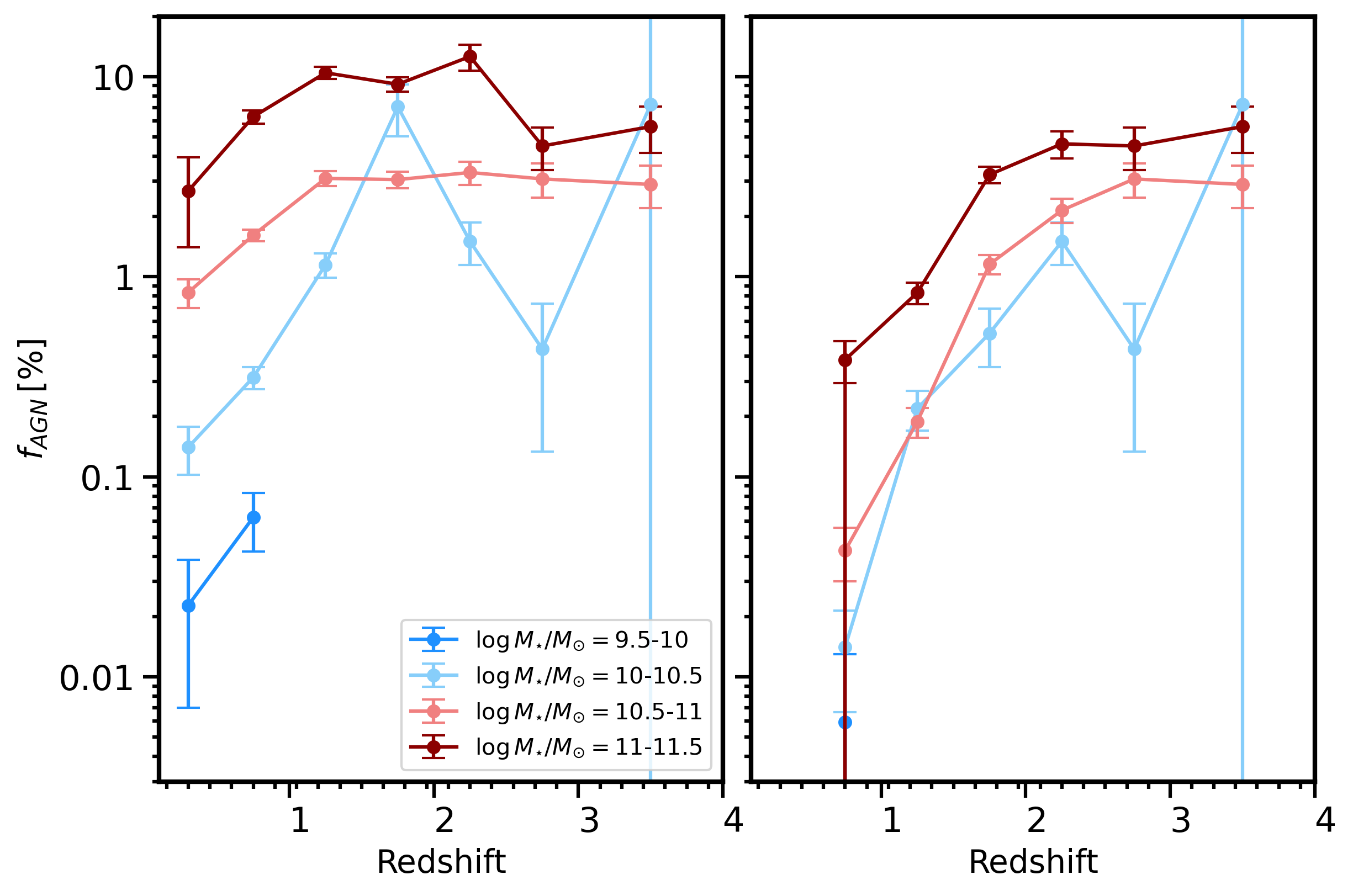}
    \caption{\textit{(Top panels)} AGN fractions to two luminosity limits in the 2D $M_{\star} - z$ plane. Brighter colour indicates a higher AGN fraction among galaxies in that bin. The left-hand panel includes X-ray detections above $L_X\geqslant \, 10^{43}\, \rm erg/s$ whereas the right-hand panel includes only X-ray bright AGN of luminosity $L_X\geqslant \, 10^{44}\, \rm erg/s$. Red dotted lines are the mass completeness limit. \textit{(Bottom panels)} Projected AGN fraction onto the redshift axis in different stellar mass bins. We keep the same luminosity cuts on both panels.}
    \label{fagn_2D_LX}
\end{figure}
To account for sensitivity effects, we correct our measurements for X-ray completeness. This correction lifts the observational bias towards higher luminosity or lower redshift AGN at fixed stellar mass, which is crucial to accurately measure the relation between AGN activity and the stellar mass of galaxies.

From the \textit{Chandra} sensitivity curve in Boötes \citep[see][and Section \ref{sec:data_X}]{masini_chandra_2020}, we infer the probability of detecting an X-ray source at a given flux limit, $p_{det}(f_X)$, by dividing the area curve by $\rm 9.3\, deg^2$, the total area of the survey. To realistically reflect the number of X-ray AGN within a given galaxy sample, we need to compensate our number of detections for the sources we miss due to sensitivity limitations. This is done using two different methods based on the detection probability, $p_{det}(f_X(L_X,z))$, where we indicate how the X-ray flux is a function of both $L_X$ and $z$. 
Firstly, we up-weight each individual detection by $1/p_{det}$ to compensate for the fact that fainter sources will be under-represented in the detected sample. Thus, our estimate of the fraction of galaxies hosting an X-ray AGN with luminosity $\geq L_\mathrm{thresh}$ is given by 
\begin{eqnarray}
    f_{AGN}(L_X \geq L_\mathrm{thresh}) = \frac{\sum^{N_{det}}_i\frac{1}{p_{det}(f_X(L_i,z_i))}}{N_{gal}} ,
    \label{eq:nx_corr}
\end{eqnarray}
where $N_{gal}$ is the total number of galaxies in a $M_{\star} - z$ bin. $L_i$ and $z_i$ are the observed luminosity and redshift of each X-ray detection and the summation is performed over the $N_{det}$ X-ray detections with $L_i>L_\mathrm{thresh}$.
Alternatively, we can reduce the effective number of \emph{galaxies} according to the number where we expect to have sufficient sensitivity to detect an AGN of luminosity $>L_\mathrm{thresh}$, and thus estimate the AGN fraction as
\begin{eqnarray}
    f_{AGN}(L_X \geq L_\mathrm{thresh}) = \frac{N_{det}}{\sum^{N_{gal}}_j p_{det}(f_X(L_\mathrm{thresh},z_j))} ,
    \label{eq:ngal_corr}
\end{eqnarray}
where the summation is now performed over the number of galaxies in a $M_{\star} - z$ bin, which are weighted according to the probability of detecting a source at the limiting luminosity, converted to a flux according to the observed redshift, $z_j$ of each galaxy.

Errors on $f_{AGN}$ are computed based on the Poisson error in $N_{det}$ following the procedure described in \cite{gehrels_confidence_1986}. For clarity, we present in Figures \ref{fagn_2D_LX} and \ref{fig:2Dfagn} our corrected AGN fractions using the first approach only, and compare both methods in Figure \ref{pledd} as they are distinct calculations that may produce different measurements.

\subsubsection{$f_{AGN}$ to different X-ray luminosity limits as a function of stellar mass and redshift}
\label{sec_fagn_LX}

The top panels of Figure \ref{fagn_2D_LX} shows the completeness-corrected AGN fraction in the 2D stellar mass - redshift space, with darker hues indicating a higher detection rate. X-ray luminosity can be used as a tracer for BH growth through accretion, so we apply a luminosity cut of $L_X \rm \, \geqslant \, 10^{43}$ and $10^{44}\, \rm erg/s$ in the left and right panels, respectively, as a way to identify moderately and highly accreting BHs. We created 15 evenly spaced mass bins and size-dependent redshift bins with a minimum of 60,000 galaxies per redshift bin (across all $M_{\star}$). Only $M_{\star} - z$ bins with $N_{count} \rm \geqslant 500$ are plotted. In both panels, we see a clear mass limit of $M_{\star} \sim \, 10^{10}\, \rm M_{\odot}$ below which the AGN fraction is negligible, close to $f_{AGN} \sim 0.5\%$. This reaffirms our prior results, that X-ray selected AGN are rarely found in galaxies with lower stellar masses. While the AGN fraction remains relatively constant at fixed stellar mass across redshifts for the $\log L_X\geqslant\, 43\, \rm erg/s$ galaxy population (left panel of Figure~\ref{fagn_2D_LX}), there is a stronger increase with redshift in the fraction of galaxies with high-to-moderate stellar masses ($M_{\star}\gtrsim10^{10.5}\, \rm M_\odot$) that host a high-luminosity X-ray AGN. 

These trends can be seen more clearly when projecting the AGN fraction onto the redshift axis and using colours to indicate the different stellar mass ranges (bottom panels of Figure \ref{fagn_2D_LX}). We find a strong increase in $f_{AGN}$ towards higher redshift for high-luminosity AGN (bottom right panel of Figure \ref{fagn_2D_LX}), increasing from $\sim 0.02$\% to $\sim3$\% in galaxies of $M_{\star}  = 10^{10-11}\, \rm M_{\odot}$ and reaching an even higher fraction of $\sim10\%$ in our highest mass bin by $z=3-4$. This strong redshift dependence almost completely vanishes for more moderately accreting BHs (\textit{left panel)}), showing only an AGN fraction increase in lower mass galaxies $M_{\star}  \leqslant 10^{10.5}\, \rm M_{\odot}$ at $z \leqslant 1.5$. This suggests that rapid BH growth predominantly occurred at higher redshifts, which reflects the evolution of the X-ray Luminosity Function (XLF), i.e. AGN of high luminosity were more common at earlier cosmic times \citep{aird_evolution_2015}.

When looking at fixed redshift, the AGN fraction appears to increase with stellar mass, especially for AGN with $L_X \geqslant 10^{43}\, \rm erg/s$. However, this could be an effect of selecting accreting BHs based on their X-ray luminosity. Indeed, the luminosity might not provide reliable insights on the accretion mechanisms at play as it can be biased by the stellar mass of the galaxy, producing similar $L_X$ for a broad range of accretion rates depending on its mass.
To account for this possible degeneracy, in the following sections we instead use the specific BH accretion rate, $\lambda$, as a more reliable tracer for BH accretion.

\begin{figure}
    \centering
    \includegraphics[width=\columnwidth]{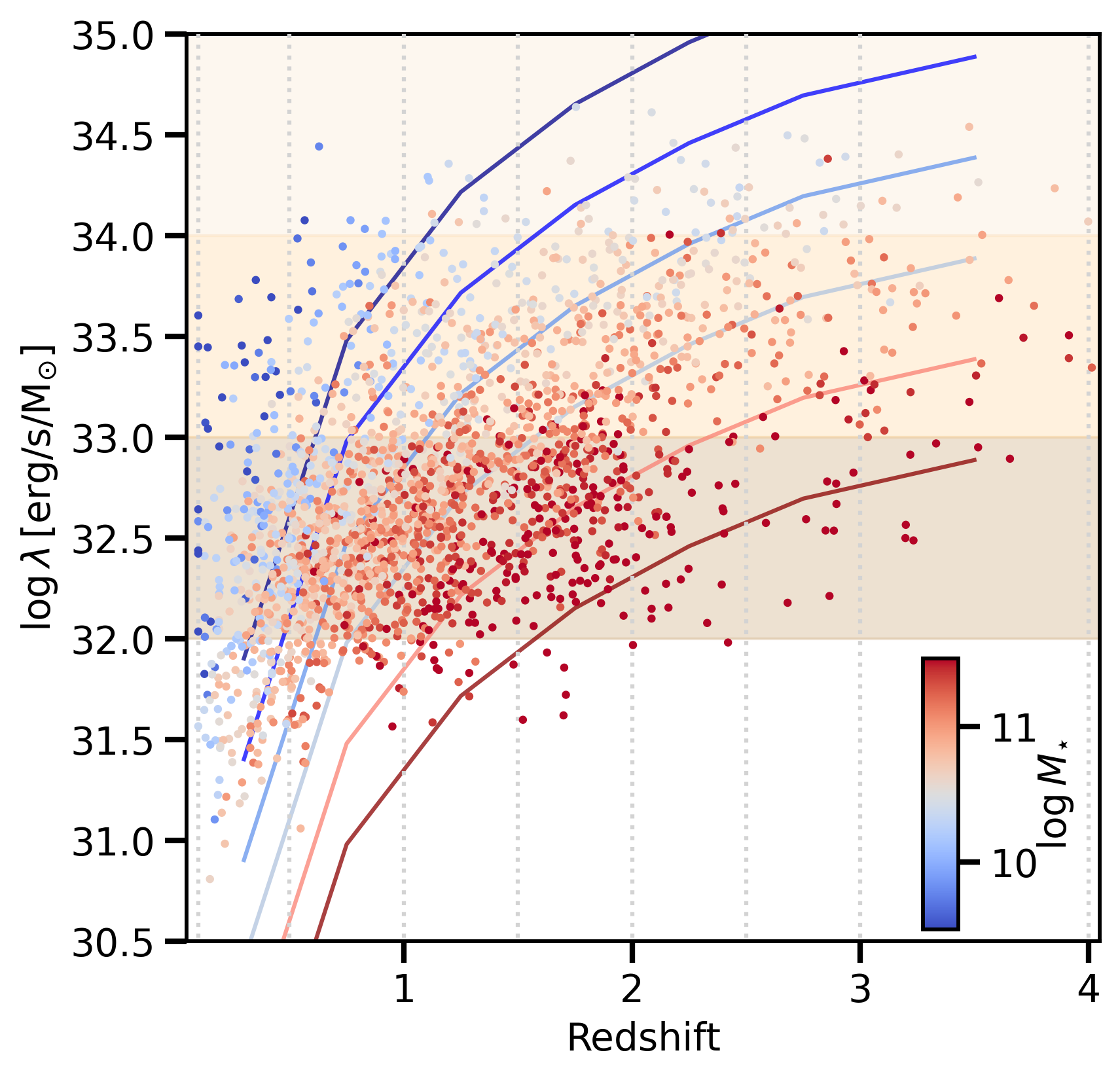}
    \caption{Distribution of stellar masses of the X-ray detected galaxies in the $\lambda - z$ space. The coloured lines show the sensitivity limits at fixed stellar mass (as indicated by the colour bar) above which 90$\%$ of sources are detected. The sensitivity limitations become more important at lower stellar mass, where only the highest-$\lambda$ AGN can be detected, especially at $z>1$, where we don't detect any low-mass galaxies ($M_\star \leqslant 10^9 \, \rm M_\odot$) below $\log \lambda = 34 \, \rm erg/s/M_\odot$. Dotted vertical grey lines show our redshift bins. Finally, the brown shaded areas show the $\lambda$ limits of $\log \lambda > 32,33,34 \, \rm erg/s/M_\odot$ as per Figure \ref{fig:2Dfagn}. We mostly find low to moderate mass galaxies in the highest $\lambda$ region.}
    \label{fig:sbhar_z}
\end{figure}

\subsubsection{Specific BH accretion rate distribution}

\label{sec:extract}

\begin{figure*}
    \centering
    \includegraphics[scale=0.67]{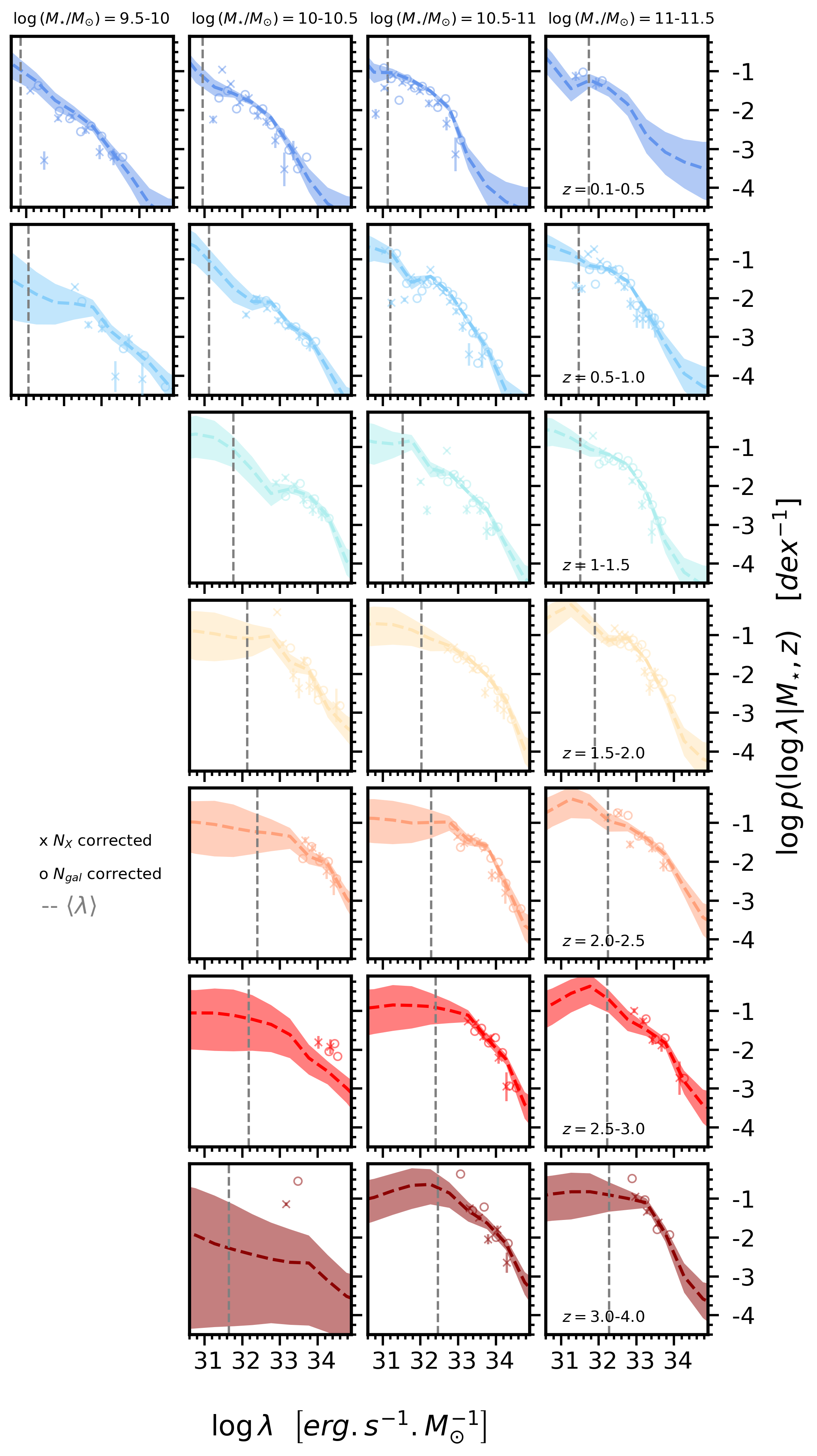}
    \caption{Probability density function of sBHAR, $\lambda$, in bins of stellar mass and redshift. Stellar mass increases to the right and redshift downward. The dashed coloured lines show the intrinsic PDF retrieved through the Bayesian modelling method described in \citet{aird_x-rays_2018} (see Section \ref{sec:extract}) and the shaded areas are the 3$\sigma$ uncertainties. We compare the model to our previous binned estimates corrected for survey completeness. The crosses are the $N_X$-corrected distributions (Equation \ref{eq:nx_corr}) and the circles are the $N_{gal}$-corrected distributions (Equation \ref{eq:ngal_corr}), all three showing good agreement, with small discrepancies in the low accretion rate regime. The dotted grey lines show the mean sBHAR $\lambda$ increasing both with redshift and stellar mass.}
    \label{pledd}
\end{figure*}

\begin{figure*}
    \centering
    \includegraphics[scale=0.65]{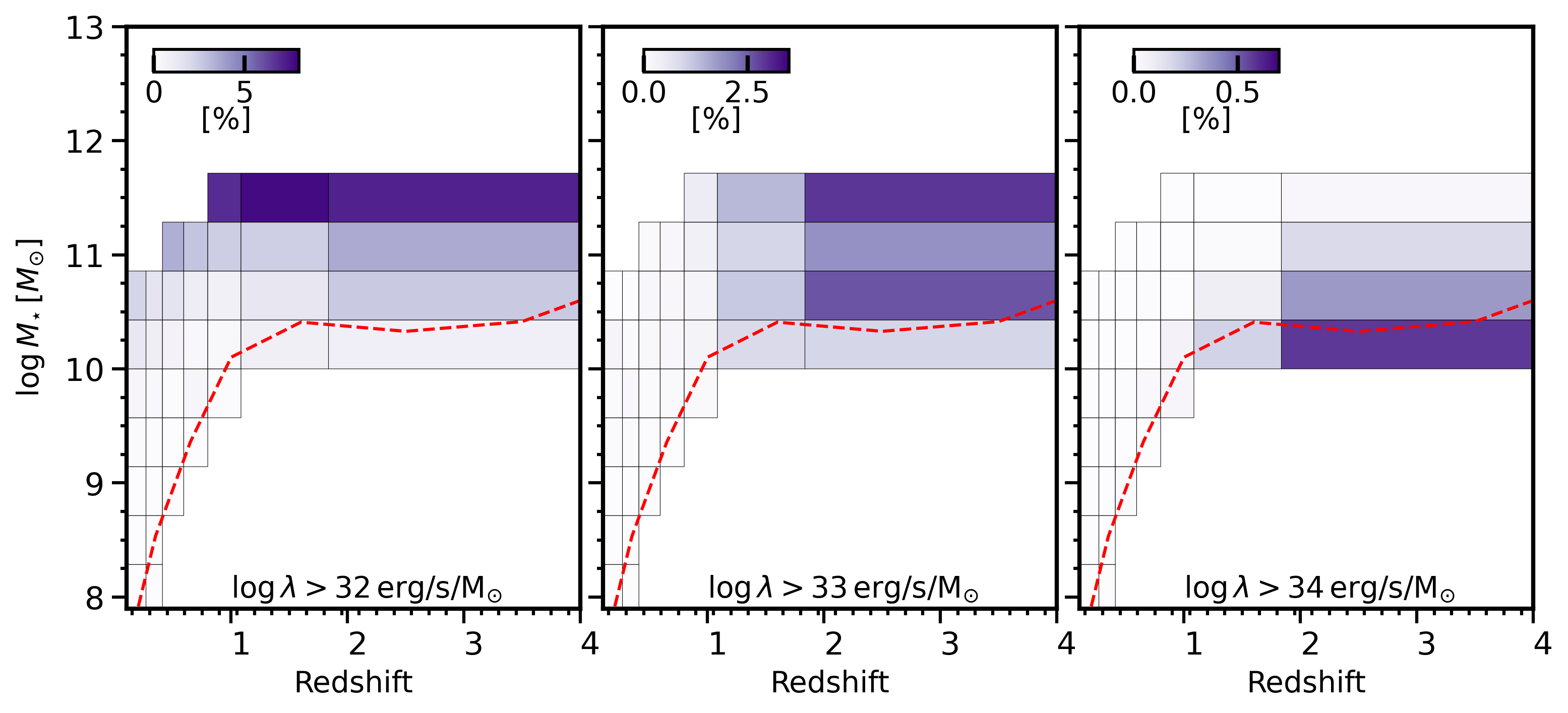}
    \caption{AGN fraction to different specific BH accretion rate limits for galaxies in the $M_{\star}-z$ plane. Brighter colour indicates a higher AGN fraction in that bin. We apply on each panel a sBHAR limit of $\log \lambda$ [erg~s$^{-1}$~$\rm M_{\odot}^{-1}$] $> 32$, 33, and 34 increasing to the right.}
    \label{fig:2Dfagn}
\end{figure*}

\begin{figure*}
    \centering
    \includegraphics[scale=0.65]{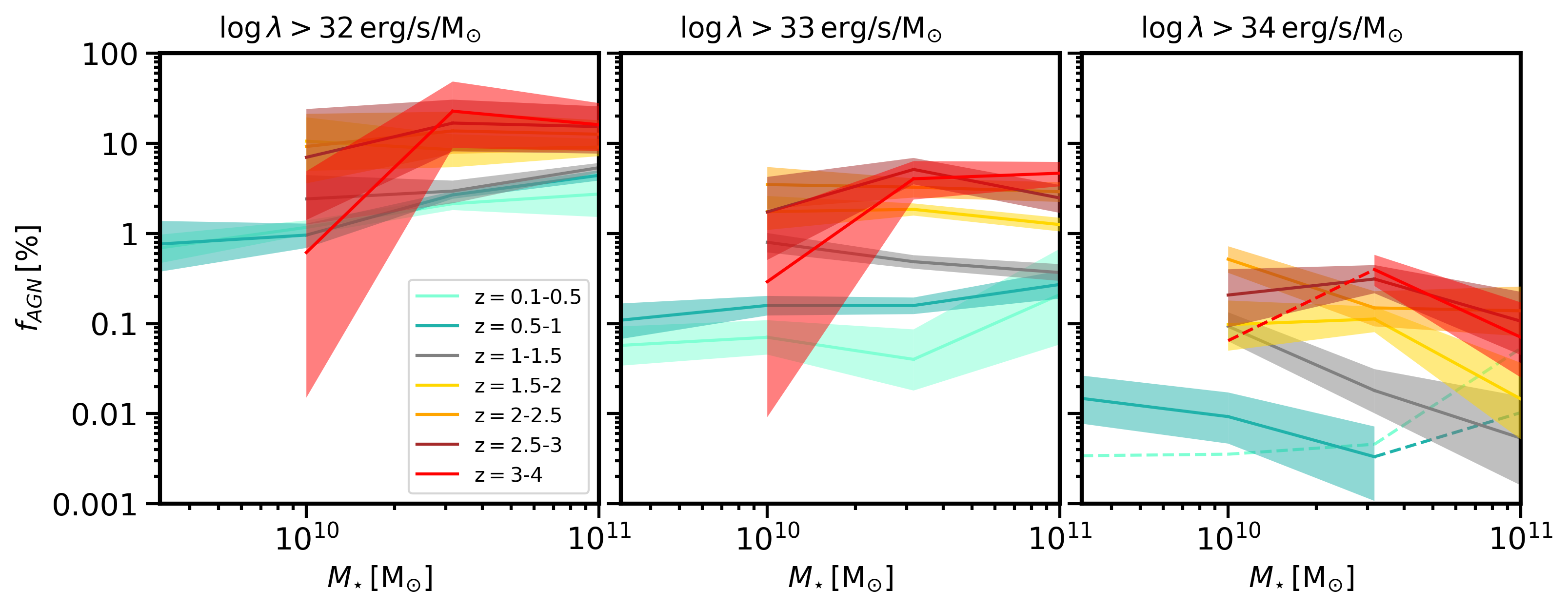}
    \includegraphics[scale=0.65]{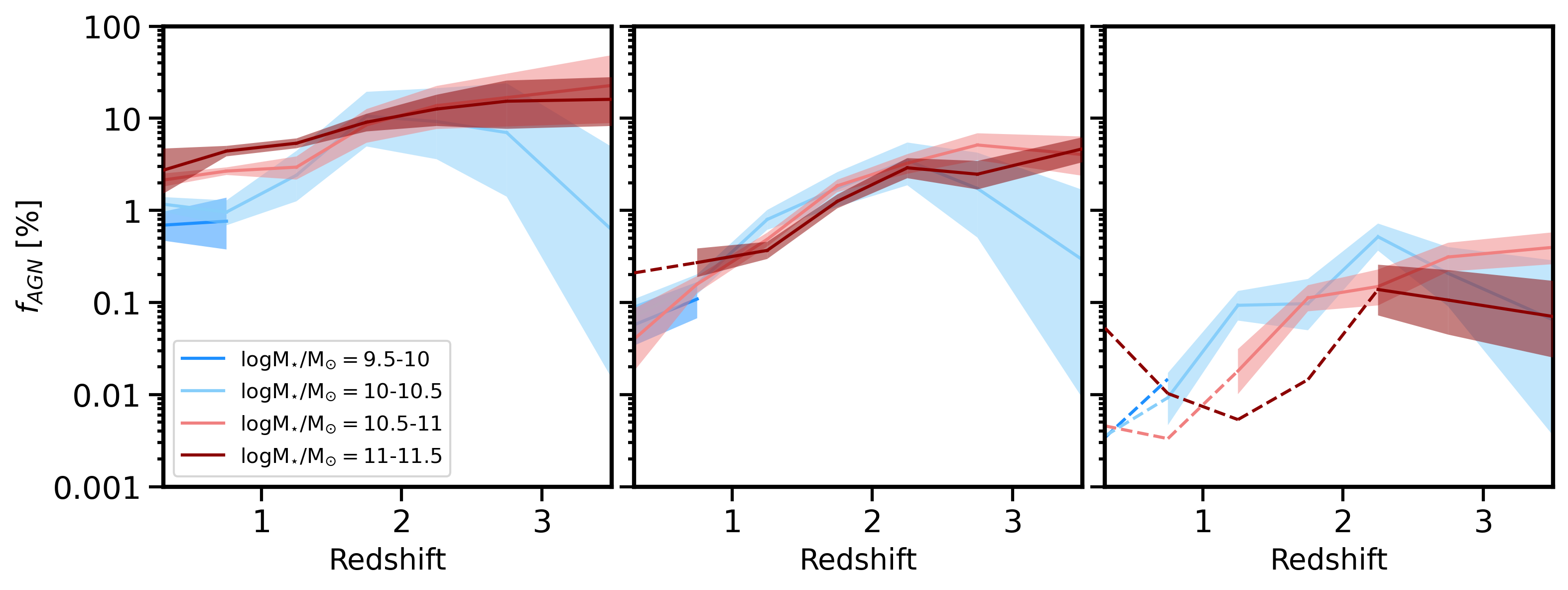}
    \caption{ \textit{(Top)}. AGN fraction evolution with stellar mass in different redshift bin. We apply on each panel a sBHAR limit of $\log\lambda > 32-33-34\, \rm erg/s/M_{\odot}$ increasing to the right. \textit{(Bottom)}. AGN fraction evolution with redshift in different stellar mass bins. sBHAR limits are the same as above. $f_{AGN}$ was computed by integrating our modelled $p(\lambda |M_{\star},z)$ described in Section \ref{sec:extract} as per Equation \ref{eq:fagn}. Dashed parts of the lines show where we don't have sources in a $M_\star - z - \lambda$ bin, and where $p(\lambda |M_{\star},z)$  was obtained by extrapolating from lower-$\lambda$ sources.}
    \label{fagn_m}
\end{figure*}

We define the specific black hole accretion rate (sBHAR) in terms of the X-ray luminosity relative to the galaxy stellar mass as in \cite{bongiorno_accreting_2012, aird_agn_2022}, i.e.

\begin{eqnarray}
        \lambda = \frac{L_{(2-10)\rm keV}}{M_{\star}}.
    \label{eq:sbhar}
\end{eqnarray}

We choose here to use these \emph{observed} quantities for clarity and to avoid implying any assumption on bolometric correction or the scaling between galaxy stellar mass and central black hole mass. Making these assumptions would allow one to convert the sBHAR into a ratio relative to the Eddington limit ($L_\mathrm{Edd}$), i.e.

\begin{eqnarray}
    \lambda_\mathrm{Edd} = \frac{L_\mathrm{bol}}{L_\mathrm{Edd}}
    \approx \frac{k_{bol}L_X}{1.3\times 10^{38}\times 0.002M_{\star}}
    \approx 10^{-34} \times \lambda,
\end{eqnarray}
with $k_{bol}$ the bolometric correction factor commonly set at $k_{bol}=25$ and the factor 0.002 represents a fiducial scaling between BH mass and host galaxy stellar mass, resulting in the final approximate translation between our sBHAR, $\lambda$, and the Eddington ratio, $\lambda_\mathrm{Edd}$.

Taking $\lambda$ from equation~\ref{eq:sbhar} as a tracer for accretion rather than $L_{X}$ lifts the degeneracy between weakly accreting BHs in higher mass galaxies and highly accreting BHs in lower mass galaxies. Indeed, we find that AGN can exhibit similar X-ray luminosities despite vast differences in their accretion rates, depending on the stellar mass of their galaxy. A BH with low accretion rate in a massive galaxy could appear as bright as a high accretion rate BH in a less massive galaxy \citep{bongiorno_accreting_2012,bongiorno_agn_2016,aird_x-rays_2018}. In Figure \ref{fig:sbhar_z}, we show the range in specific accretion rate $\lambda$ we are covering across redshift. Individual X-ray sources are coloured by their stellar mass, showing a gradient of decreasing stellar mass with increasing $\lambda$. Similarly to an X-ray luminosity detection limit, we compute the survey sensitivity limits in terms of sBHAR at fixed stellar mass, shown by the solid coloured lines. We retrieve the flux limit value at $p_{det} = 0.1$ from the area curve of the survey so that $90\%$ of the X-ray sample is contained above this flux limit, we then turn it into an X-ray luminosity at the low end of each redshift bin and finally compute an sBHAR limit at the low end of our stellar mass bins. These $\lambda$ limits indicate that in the lowest mass galaxies we are only able to detect the highest $\lambda$ sources.

In this section, we present the intrinsic sBHAR probability distribution $p(\log \lambda | M_{\star},z)$ defined in the $\log \lambda$ space, to characterise in more detail BH growth in our sample across redshift and stellar mass. This quantity represents the probability for a galaxy to harbor a BH accreting at a rate $\lambda$, given its mass and redshift ($M_{\star}, z$).

In addition to the X-ray completeness corrections shown previously in Section \ref{sec:acurve}, we present here a more sophisticated approach to retrieve an unbiased intrinsic sBHAR probability distribution, similar to the approach taken by \cite{aird_x-rays_2018} where they use a non-parametric model to  recover $p(\log \lambda | M_{\star},z)$. We extract X-ray data at the position of every galaxy in our sample so as to include the non-detected sources (see Section \ref{sec:data_X}). This allows us to push the sensitivity below the X-ray detection limit and thus apply new constraints in the low accretion rate regime. We assume that the X-ray emission is always due to an AGN and do not correct for the contribution of X-ray binaries or extended X-ray emitting gas, which is a reasonable assumption given the relative depths of our X-ray imaging. We convert counts into rest-frame luminosities in the $2-10$~keV hard band, assuming a spectral index $\Gamma=1.9$, and divide by the measured stellar mass to recover sBHARs. 
The range of possible accretion rates for both X-ray detections and non-detections are modeled by Poisson likelihoods \citep[as in][]{aird_x-rays_2017,aird_x-rays_2018}:

\begin{eqnarray}
    \mathcal{L}(N_i | \lambda, b_i, t_i, z_i) = \frac{(k_i\lambda + b_i)^{N_i}}{N_i !} e^{-(k_i\lambda + b_i)}
\end{eqnarray}
which are used to place constraints on a model for the intrinsic distribution of accretion rates, $p(\log \lambda | M_{\star},z)$. $N_i, b_i, t_i, z_i$ are the total X-ray counts, background count rate, exposure time and redshift of the $i$-th source. $k_i$ is the factor depending on $t_i$ to convert sBHAR into the observed count rate. Here, we adopt a step function parametrization of $p(\log \lambda | M_{\star},z)$ over a discrete grid in $\log \lambda$ \citep[in contrast to][who model the intrinsic distribution as a series of Gamma functions, which is equivalent to using a fine enough $\lambda$ grid]{aird_x-rays_2017,aird_x-rays_2018}.
The combination of the Poisson distribution for each source in a bin and this step function  $p(\log \lambda | M_{\star},z)$ produces an overall likelihood function such that:

\begin{eqnarray}
    \mathcal{L}_{M_\star , z} = \prod_{i=1}^{N_{source}} \int_{-\infty}^\infty \mathcal{L}(N_i | \lambda, b_i, t_i, z_i) \times p(\log \lambda | M_{\star},z) \mathrm{d}\log \lambda
\end{eqnarray}

This function can be reduced to a series of weights from each source (whether detected or not) multiplied by the normalization of each step as described by \cite{aird_x-rays_2017,aird_x-rays_2018}.
Finally, a Gaussian smoothness prior $\sigma = 0.6$ is applied to ensure smooth variation in $p(\log \lambda | M_{\star},z)$  over adjacent steps in the $\log \lambda$ grid. We fitted this model using the \ascii{STAN} code \citep{Stan} which explores the parameter space with MCMC sampling and converges using maximum likelihood estimation.

In Figure \ref{pledd}, we show the corrected $p(\log \lambda |M_{\star},z)$ obtained using the three methods described in this paper: the two methods of correcting for completeness of the X-ray detected sample described in Section \ref{sec:acurve} and the non-parametric modelling that uses the data from both X-ray detections and non-detections described above. The stellar mass bins increase to the right and redshift bins increase from top to bottom. The crosses show the binned distributions correcting $N_X$ as per Equation \ref{eq:nx_corr}, while the circles are the $N_{gal}$ corrected distributions following Equation \ref{eq:ngal_corr}. Shaded areas and dashed lines are the results of the non-parametric modelling. The vertical grey dotted lines indicate the average sBHAR over all galaxies per stellar mass-redshift bin computed across a fixed $\lambda$ grid: 
\begin{eqnarray}
    \langle \lambda(M_{\star},z)\rangle = \int \lambda p(\log \lambda |M_{\star},z)\mathrm{d}\log\lambda.
    \label{eq:mean_lambda}
\end{eqnarray}

For low-redshift galaxies ($z \sim 0.1 - 1$) we identify a break at $\log\lambda \lesssim 32\, \rm erg/s/M_{\odot}$ ($1\%$ Eddington), slightly increasing with stellar mass, indicating the lack of AGN with the highest BH growth rates at later times, possibly due to the relative inefficiency of the processes that drive gas toward the central BH in the local universe \citep{kauffmann_unified_2000}. At higher redshifts, we observe a shift in the $\lambda$-break towards higher accretion rate, reaching a maximum at $\log\lambda \sim 34.2\, \rm erg/s/M_{\odot}$ (slightly above the expected Eddington limit) in galaxies of $M_{\star} = 10^{10-10.5}\, \rm M_{\odot}$ at $z=3$, indicating stronger accretion events occurring at higher redshift and may drive rapid BH growth. The slight stellar mass dependence observed in low-redshift galaxies now mostly vanishes at $z\geqslant 1$, and we find similar $p(\log \lambda |M_{\star},z)$ across all masses at fixed redshift, covering the same $\lambda$ range.

Overall, our three methods to derive completeness-corrected and intrinsic distributions are in good agreement with one another showing broad $\lambda$ distributions across the whole galaxy population, with small discrepancies in the low sBHAR regime where our survey is too shallow to efficiently constrain the incidence of the faintest AGN. Indeed, our modelled $p(\log \lambda |M_{\star},z)$ show a flattened distribution at low accretion rate where our sample hits a lower limit for the incidence of radiatively efficient accretion, and presenting little evolution of the average sBHAR $\langle\lambda\rangle$ with stellar mass. However, since our binned estimates measured on the corrected observed samples cannot recover the intrinsic distributions down to lower accretion rate, small discrepancies arise in that regime. Indeed, in certain cases, our binned estimates present a stronger turnover at higher accretion rate than the modelled $p(\log \lambda |M_{\star},z)$ which uses extracted X-ray data and is thus able to retrieve intrinsic distribution at lower $\lambda$, especially in the medium redshift-mass range (first yellow panel at $z=1.5-2$, $M_{\star} = 10^{10-10.5}\, \rm M_{\odot}$).
In the following, we will proceed with our corrected measurements through the non parametric modelling approach only as it proves to better recover the intrinsic specific accretion rate distribution across the whole $\lambda$ range, as opposed to our binned estimates where the correction on individual sources in the poorly constrained low-$\lambda$ regime is less reliable.

\subsubsection{$f_{AGN}$ to different specific BH accretion rate limits as a function of stellar mass and redshift}
\label{sec_fagn_sBHAR}

We can derive the AGN fraction from the sBHAR probability density function, $p(\lambda | M_{\star},z)$, such that,
\begin{eqnarray}
        f_{AGN}\left(\lambda > \lambda_{thres} | M_{\star},z\right) = \int_{\lambda_{thres}}^{\infty} p(\log \lambda | M_{\star},z) \mathrm{d}\log \lambda.
        \label{eq:fagn}
\end{eqnarray}

In Figure \ref{fig:2Dfagn}, we present an overview of our $f_{AGN}$ measurement in the stellar mass - redshift space. In each of the panels, we apply a sBHAR threshold increasing to the right from $\log\lambda \rm= 32-33-34 \, \rm erg/s/M_{\odot}$. The colour scale differs from one panel to another for better contrast between low and high AGN fractions. This configuration allows us to distinguish where the AGN population is shifting in the $M_{\star}-z$ space with increasing accretion rates. We see again that the low-redshift AGN population rarely consists of higher $\lambda$ sources, and that the peak of AGN fraction (dark purple region) gradually shifts towards higher redshift and lower stellar masses for higher accretion rates. Indeed, we observe an opposite trend of the AGN fraction with stellar mass when $\lambda$ increases. At lower $\lambda$, the peak of AGN fraction is located in the most massive galaxies for a given redshift, whereas for higher accretion rate BHs (\textit{right panel}), the AGN fraction is higher in lower mass galaxies $\lesssim 10^{11}\, \rm M_{\odot}$. Those results are shown in more detail in Figure \ref{fagn_m}, where we show the AGN fraction as a function of stellar mass for each redshift bin (\textit{top row}) and as a function of redshift for each stellar mass bin (\textit{bottom row}). In each column, we apply the same sBHAR threshold as in Figure \ref{fig:2Dfagn}. 

At a given redshift \textit{(top panels of Figure \ref{fagn_m})}, when selecting AGN to limits in terms of specific accretion rate, we see that the $f_{AGN} - M_{\star}$ relation flattens with increasing sBHAR $\lambda$ threshold, suggesting that the stellar mass dependence in the fraction of galaxies that host AGN depends on the choice of AGN selection threshold. This shows that a significant fraction of lower mass galaxies contain AGN with moderate-to-high accretion rates, and that this fraction is only weakly dependent on stellar mass. While for the lowest $\lambda$ threshold we find a slight increase in the AGN fraction with stellar mass, for higher accretion rates we find no clear predominance at a certain host mass range. For medium-high accretion rate $\log\lambda > 33\, \rm erg/s/M_{\odot}$ ($\sim 10\%$ Eddington), we find a constant $f_{AGN}$ across a wide stellar mass range of $M_{\star} = 10^{9.5-11}\, \rm M_{\odot}$ at fixed redshift, confirming our findings in Section \ref{sec_fagn_LX}. When considering more extreme accretion rates ($\log\lambda > 34\,\rm erg/s/M_{\odot}$), a mostly flat $f_{AGN} - M_{\star}$ is found at all redshifts, although we identify a drop at higher masses (for $z=0.5-2.5$), suggesting a suppression of AGN activity in the most massive galaxies at all but the most recent cosmic times. 

This same hierarchy is also apparent in the lower panels of Figure \ref{fagn_m} that show the AGN fractions as a function of redshift.
At fixed stellar mass, we find that the AGN fraction increases with redshift for all $\lambda$ thresholds as previously seen in Section \ref{sec_fagn_LX} for X-ray luminosity thresholds. 
However, for the lowest $\lambda$ threshold we find the highest AGN fractions in higher mass galaxies (at all redshifts), but this trend reverses for the higher $\lambda$ thresholds.
We also find a relatively weaker redshift dependence for low accretion BHs. For the moderate $\lambda$ threshold ($\log \lambda>33 \rm erg/s/M_{\odot}$) the AGN fraction increases by a factor of almost 100 between $z\sim0$ and $z\sim2$ for all stellar mass ranges, whereas for a lower $\lambda$ threshold the increase is much milder (a factor $\lesssim10$ over the redshift range), indicating that slowly growing BHs can be found at all redshift, with a higher fraction found in massive galaxies as shown by the dark red curve.
In contrast, very active BHs with high accretion rates are mostly found at high redshift and are located over a more diverse galaxy population, with a peak AGN fraction of around $0.5\%$ in galaxies of all masses at the highest redshifts. 

While high mass galaxies typically contain the largest fraction of AGN when considering lower accretion rates, they lack the most rapidly accreting sources which indicates a suppression in the most rapid levels of black hole growth within such galaxies.

\section{Discussion}
\label{sec:discussion}
In this paper, we have presented new measurements of the AGN incidence within massive galaxies through the evaluation of the AGN fraction $f_{AGN}$ and probability density function of accretion rates $p(\lambda|M_{\star},z)$ as a function of stellar mass and redshift, using X-ray survey data. In section \ref{sec:comparison}, we compare our results with prior literature. In Section \ref{AGN_host}, we then discuss what our results tell us about the types of galaxies where AGN are found and where most BH growth occurs. Finally, in Section \ref{sec:growth_tracks}, we assess whether this growth is sufficient to produce the very massive BHs observed in the local Universe.

\subsection{Comparison with previous studies of AGN fractions}
\label{sec:comparison}
\begin{figure*}
    \centering
    \includegraphics[scale=0.65]{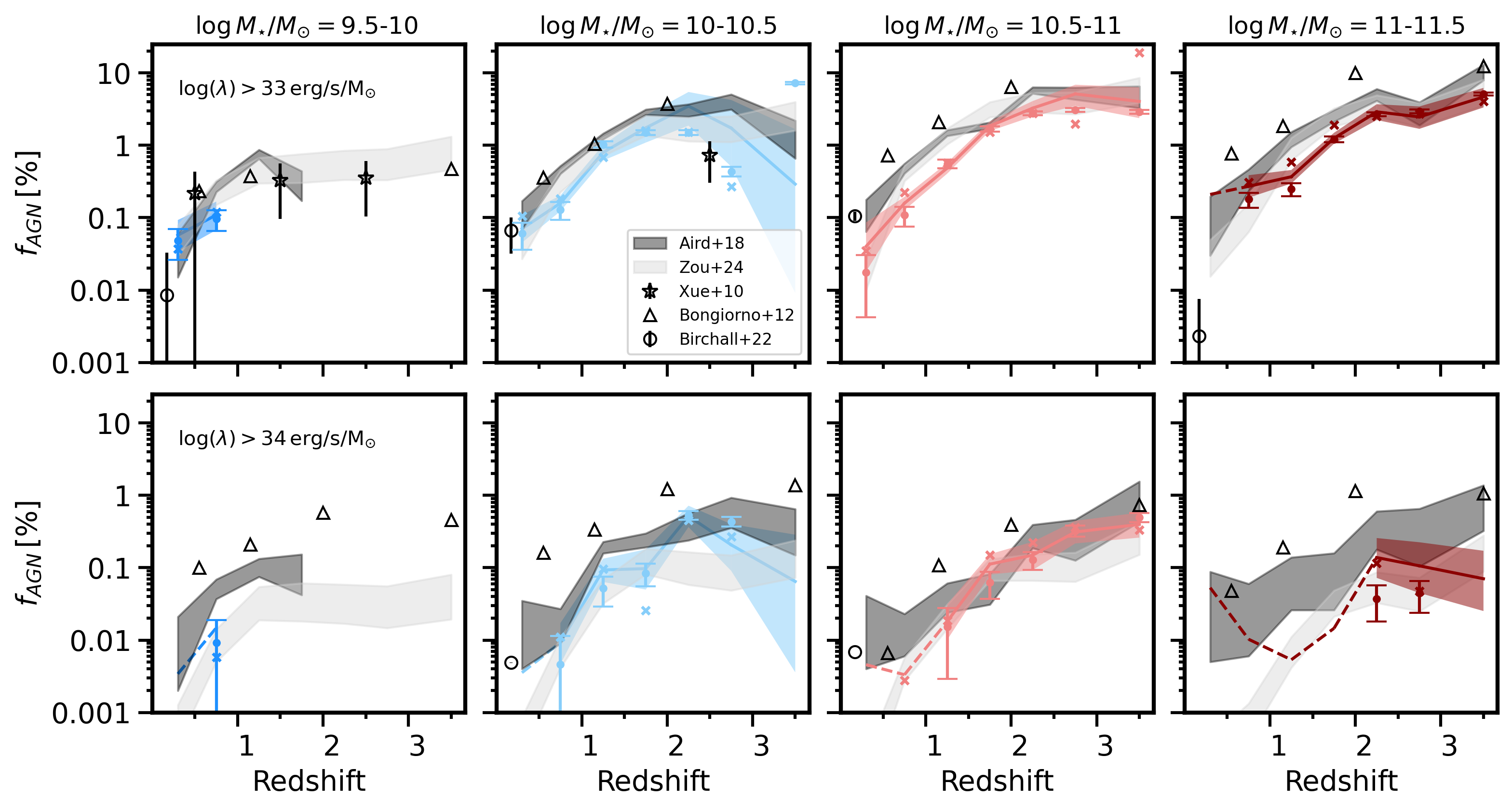}
    \caption{Comparison of $f_{AGN}-z$ relations from \citet{xue_color-magnitude_2010} (black stars), \citet{bongiorno_accreting_2012} (black triangles), \citet{birchall_incidence_2022} (black circles), \citet{aird_x-rays_2018} (black shaded regions) and \citet{zou_mapping_2024} (grey shaded regions). Our measurements are shown in colour, featuring all correction methods. The crosses are the $N_{X}$-corrected fractions (Equation \ref{eq:nx_corr}), the full circles are the $N_{gal}$-corrected fractions (Equation \ref{eq:ngal_corr}) and the shaded areas are the AGN fractions derived from our modelled distributions of sBHAR (see Section \ref{sec:extract} and Equation \ref{eq:fagn}). Our measurements are broadly consistent with prior work (see Section \ref{sec:discussion} for a detailed discussion).}
    \label{fig:fagn_comp}
\end{figure*}

\begin{figure}
    \centering
    \includegraphics[scale=0.65]{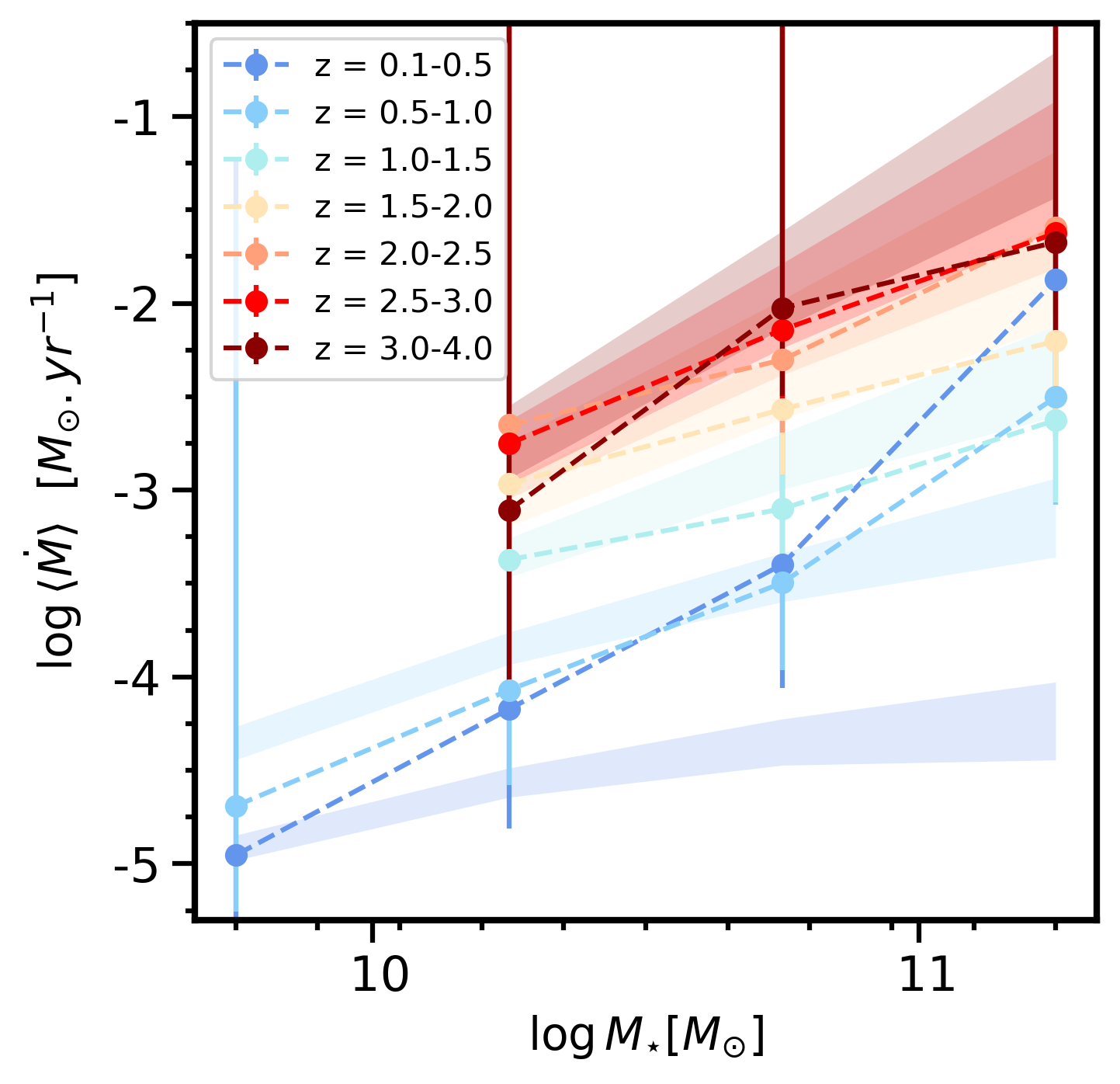}
    \caption{Average BH accretion rate computed following Equation \ref{mdot}. We compare it to the measurements of \citet{yang_linking_2018} (shaded areas).}
    \label{fig:mdot}
\end{figure}

In Figure \ref{fig:fagn_comp}, we compare our findings to various AGN fraction measurements made in prior studies using  X-ray survey data to constrain BH accretion. We focus on the high accretion rate regime, $\log \lambda\, \geqslant\, 33-34\, \rm erg/s/M_{\odot}$, where most of BH growth occurs. The majority of the works we compare with also use $\lambda$ as a tracer for accretion rate, to account for observational bias by normalizing by the stellar mass.

Earlier measurements by \cite{bongiorno_accreting_2012} (black triangles) are mostly consistent with our measurements across cosmic time and stellar mass, finding negligible stellar mass dependency of the AGN fraction at a given redshift. Their measurements were based on the $\sim 40$ks \textit{XMM-Newton} imaging of the $\sim 2$~deg$^2$ COSMOS field, whereas our wider area coverage ensures our sample contains significantly more high-mass galaxies and our deeper \textit{Chandra} data allows us to probe fainter AGN. We thus find the largest discrepancies in this massive regime of $M_{\star} = 10^{11-11.5}\, \rm M_{\odot}$, for both $\log \lambda = 33, 34 \, \rm  erg/s/M_{\odot}$ corresponding to Eddington ratios of $10\, \%$ and $100\, \%$ respectively. Although the evolutionary behaviour of the AGN fraction with redshift is similar, we systematically find a lower fraction in the highest mass galaxies than \cite{bongiorno_accreting_2012}. 

Indeed, when comparing to more recent studies, we observe closer similarities with our measurements over the entire $M_{\star} - z$ space. Probing smaller and deeper areas thanks to deep \textit{Chandra} surveys, \cite{aird_x-rays_2018} offer better constraints in the low-mass and low-accretion regime. Larger deviations arise again in the $M_{\star} = 10^{11-11.5}\, \rm M_{\odot}$ and $\log\lambda > 34\, \rm erg/s/M_{\odot}$ space, where we find on average a lower fraction but still overlapping with the range of uncertainties in their measurements. After normalizing the accretion rate by the stellar mass, they found a weak stellar mass dependence at low-redshift which we also identify in Figure \ref{fagn_m}, as well as in Figure \ref{pledd} where we only find a significant increase of average sBHAR, $\langle\lambda\rangle$, with stellar mass for $z\leqslant 1$ (dashed grey lines on Figure \ref{pledd}) where 
$\langle\lambda\rangle$ increases from $\sim 10^{31}$ to $\sim 10^{31.9} \, \rm erg/s/M_{\odot}$ for stellar masses increasing from $M_{\star} = 10^{9.5}$ to $10^{11.5}\, \rm M_{\odot}$. \citet{birchall_incidence_2022} also found little change in the AGN fraction with stellar mass in the local galaxy population (to specific accretion rate limits), identifying AGN within SDSS galaxies using X-ray data from 3XMM. While typically probing a lower redshift range than our measurements ($z=0.01-0.3$), their results (circles in Figure~\ref{fig:fagn_comp}) generally follow a consistent trend with our measurements, although we note that they measured an extremely low AGN fraction (for $\lambda>10^{33}\,\rm erg/s/M_{\odot}$) in the highest mass galaxies ($M_{\star} = 10^{11-11.5}\, \rm M_\odot$), consistent with our findings that moderate-to-high accretion rate AGN activity may be suppressed in the most massive galaxies. 
Finally, results from \cite{zou_mapping_2024} acquired in COSMOS and the wide eROSITA field, eFEDS \citep{brunner_erosita_2022}, are in good agreement with our AGN fraction measurements in the high mass and high accretion rate regime, finding a slightly lower fraction in those massive galaxies at fixed redshift. However, as they rely on eROSITA soft band observations, detections in their widest field are more sensitive to obscuration (see \cite{laloux_accretion_2024} for obscuration-corrected $\lambda$ distributions, finding an offset towards higher accretion rate for unobscured sources) which could induce a bias in fainter, low-mass galaxies. However, in the low-mass regime where \citet{zou_mapping_2024} rely on deeper fields and use hard X-ray data, the slight inconsistencies might arise from assumptions in their semi-parametric modeling of $p(\log \lambda)$ versus our flexible approach that adopts step-functions with Gaussian smoothness priors.
Such differences could explain their slightly lower AGN fraction measurements in galaxies of mass $M_{\star} = 10^{9.5-10.5}\, \rm M_{\odot}$ compared to our measurements.

Overall, our measurements agree with prior studies that found redshift-dependent AGN fractions, showing a weak dependence on stellar mass after normalization of the accretion rate, depending on where the accretion rate threshold is set. We find no evidence for a strong enhancement of the AGN prevalence in massive galaxies, measuring roughly equivalent AGN fractions in all galaxy populations at a given redshift with $f_{AGN} \sim 0.02 - 9 \%$ increasing with redshift, for all $M_{\star}$ bins at $\log\lambda > 33\, \rm erg/s/M_{\odot}$. 
In fact, our measurements suggest a slight suppression of the most rapid accretion phases ($\lambda\gtrsim 10^{34} \rm erg/s/M_{\odot}$) in galaxies of higher stellar masses. However, due to adopting a wider but shallower survey, we are not able to probe the AGN content of low-mass galaxies with $M_{\star} \leqslant10 ^{10}\, \rm M_{\odot}$ above $z\sim 1$. 

\subsection{Where in the galaxy population are X-ray AGN found?}
\label{AGN_host}

Through the measurements of stellar masses, X-ray luminosities, specific BH accretion rates and AGN fraction over the optically selected galaxy populations in Boötes, we have assessed the AGN incidence as a function of galaxy properties and studied the connection between host stellar mass and the presence of X-ray AGN.

Figure \ref{fig:M-z} shows that the host galaxies of our X-ray detected AGN sample typically have moderate-to-high stellar masses at all redshifts (median $M_\star\sim10^{10.5-11}\, \rm M_\odot$), albeit with a large spread, and are thus on average more massive than our optically-selected galaxy sample where no AGN signature was detected. However, when supplementing our analysis with intrinsic AGN fraction and specific BH accretion rate distribution across stellar mass and redshift (Figures \ref{pledd} - \ref{fig:fagn_comp}), we find that this AGN prevalence in massive galaxies is primarily the result of selection effects impacting the observed samples \citep{aird_primus_2012,aird_primus_2013}. 
When considering the fraction of galaxies that have AGN (to sBHAR limits), only a mild increase with stellar mass is found for the lowest accretion rate limits ($\log\lambda > 32 \, \rm erg/s/M_{\odot}$) and no overall mass-dependency is found for BHs accreting at rates $\log\lambda > 33 \, \rm erg/s/M_{\odot}$. It was rather a degeneracy effect driving this increase when considering X-ray luminosity. 
For BHs accreting at Eddington rates and above ($\log\lambda > 34 \, \rm erg/s/M_{\odot}$), we find a drop in the AGN fraction at the highest stellar masses $M_{\star} \geqslant \, 10^{10.5}\, \rm M_{\odot}$, indicating a suppression of the highest instantaneous rates of BH growth in such galaxies. However, AGN activity remains high in these massive galaxies even below $z\sim 1$, showed by the upturn at high masses in the $z = 0.1-1$ curves of Figure \ref{fagn_m}, indicating that massive local galaxies may continue to assemble their BHs through accretion although the majority of such galaxies are already quenched at $z<1$ with highly suppressed SFRs and little ongoing BH growth.

Thus, our study reveals that AGN are indeed found across the whole galaxy population but observed samples are dominated by moderate to high mass galaxies, where deeper optical/IR surveys could reveal a substantial amount of non-active lower-mass galaxies.

Finally, Figure \ref{Mstar-LX} reveals that most BH growth in the observed, and thus flux-limited sample, occurs in moderate mass galaxies of $M_{\star} \sim 10^{10.5}\, \rm M_{\odot}$ in line with the results shown in \cite{aird_primus_2013}. This predominance of galaxies of such masses in observed samples of AGN is a consequence of the characteristic break in the galaxy stellar mass function around $M_{\star} \sim \, 10^{10.5}\, \rm M_{\odot}$. At fixed luminosity, lower mass hosts would require a much larger sBHAR AGN to produce similar luminosities, but these less massive galaxies cannot dominate the observed samples as high-$\lambda$ AGN are much rarer in low-mass galaxies. This means that at lower X-ray luminosities, flux-limited AGN samples must be dominated by a broad range of low-to-moderate specific accretion rates in massive galaxies.
While BHs seem to be growing across the full mass range of galaxies, we will assess whether this assembly is sufficient or not to account for the massive BH population found at $z\sim 0$.

\subsection{Quantifying the amount of BH growth across cosmic time}

\label{sec:growth_tracks}
\begin{figure}
    \centering
    \includegraphics[scale=0.4]{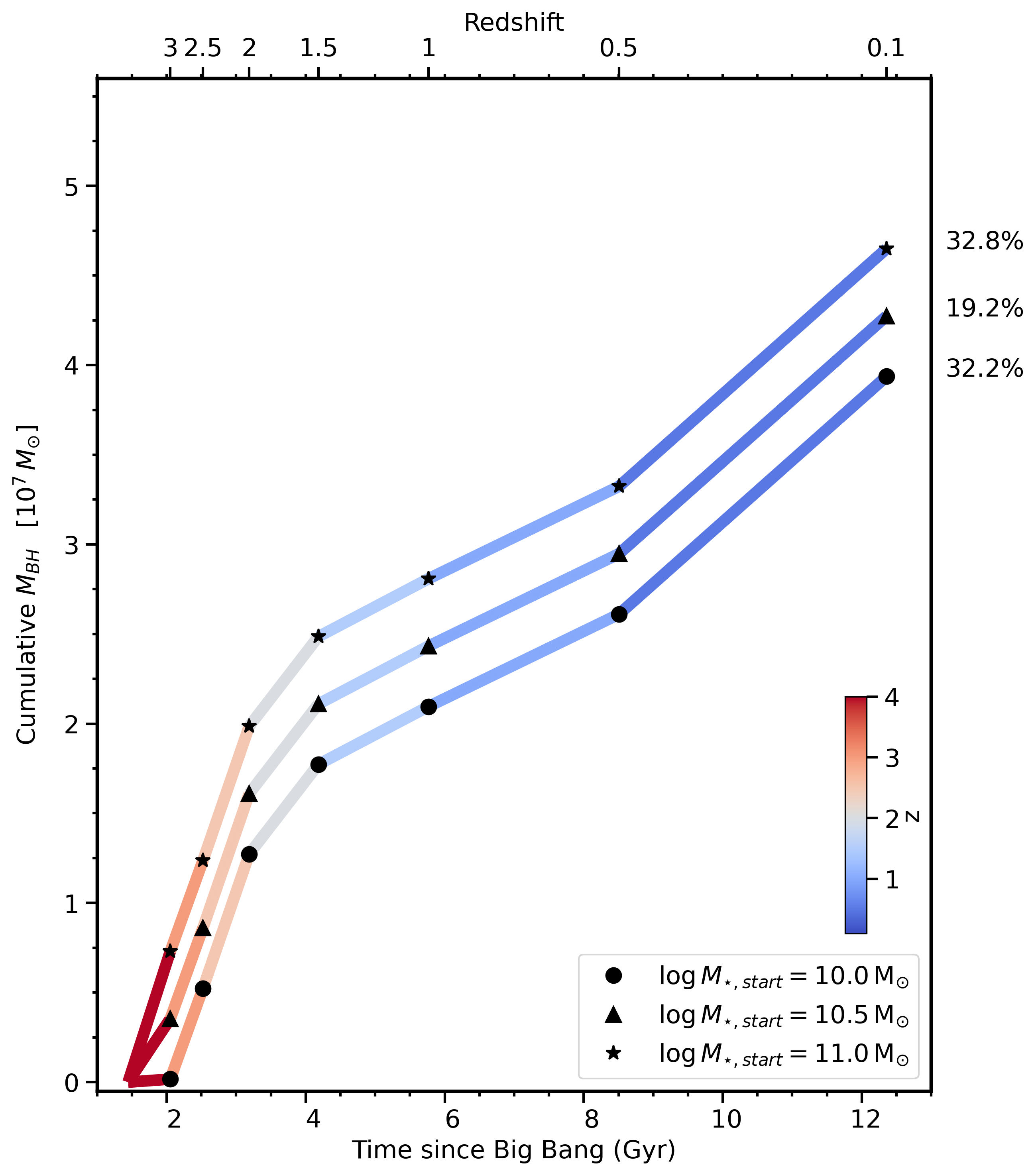}
    \caption{Cumulative BH mass accreted from redshift $z=4$ to $z=0.1$. The mass gained at each time step is computed following Equation \ref{bhmass_gain}. We show the average BH mass evolution in three different galaxies of mass $\log M_{\star,start} = 10-10.5-11\, \rm M_{\odot}$. The fractions on the side represent the ratio between the total mass grown between $z=4-0.1$ and the expected BH mass at $z=0$ according to the \citet{greene_intermediate-mass_2020} late-type scaling relation. Parallel to this process, we also grow galaxies in stellar mass, thus the expected BH masses at $z=0$ are taken at higher $M_{\star}$ than $M_{\star,start}$.}
    \label{fig:cml_mbh}
\end{figure}

\begin{figure*}
    \centering
    \includegraphics[scale=0.65]{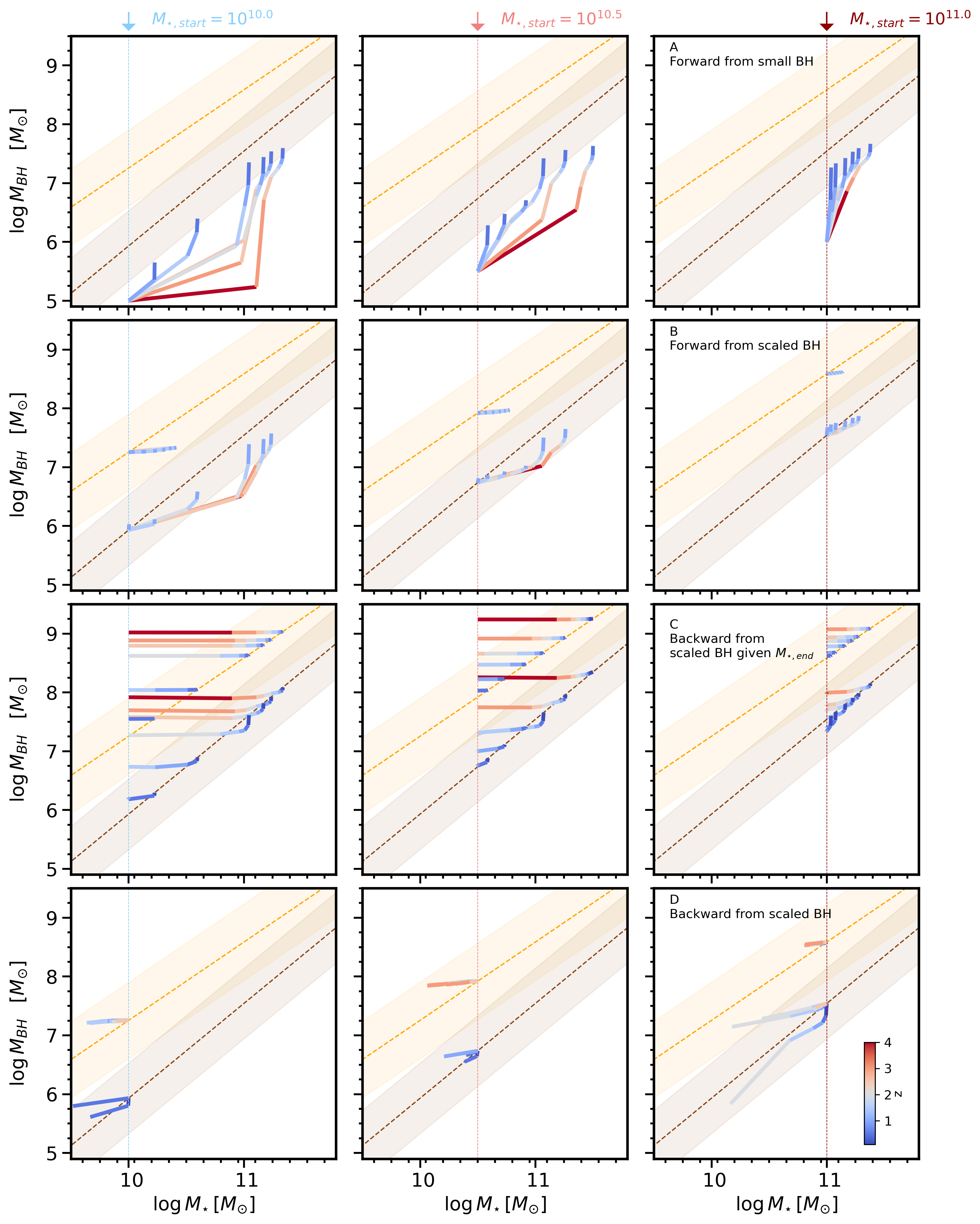}
    \caption{Black Hole - galaxy growth tracks across cosmic time for different initial stellar masses ($\log M_{\star,start} = 10-10.5-11\, \rm M_{\odot}$). Each point of a track is the total mass accumulated up to redshift $z=3,2.5,2,1.5,1,0.5,0.1$ with $z=0.1$ being the end point to the right of each track. Two sets of tracks are calculated: the \textit{forward} tracks on the top panels where we increase BH mass based on our measured accretion rates (starting from either low-mass seeds for method A, first row, or from the scaling relations for method B, second row); and the \textit{backward} tracks shown on the third and fourth rows (method C and D) where we assume a BH mass at z=0 and track the corresponding mass back in time based on our accretion rate measurements. Scaling relations from \citet{greene_intermediate-mass_2020} are shown by the orange (early-type) and brown region (late-type).
    }
    \label{fig:tracks}
\end{figure*}
In this section, we use our measurements to estimate the average amount of BH and galaxy growth to track their common assembly and growth pathways through cosmic time. 
 
The average BH accretion rate $\dot{M}$ is computed from the mean sBHAR $\langle\lambda (M_\star ,z)\rangle$. We update Equation \ref{eq:mean_lambda} (dotted grey line in Figure \ref{pledd}), to include an X-ray luminosity-dependent bolometric correction $k_{bol}$ as per \cite{lusso_bolometric_2012, yang_linking_2018} such that,

\begin{eqnarray}
    \langle \dot{M}(M_{\star},z) \rangle = \int_{28}^{\infty} p(\log\lambda|M_\star, z) \frac{1 - \epsilon}{\epsilon} \frac{k_{bol}(M_\star \lambda ) M_{\star} \lambda}{c^2} \mathrm{d}\log\lambda
    \label{mdot}
\end{eqnarray}
where we fix the radiative efficiency $\epsilon = 0.1$  \citep{marconi_local_2004, merloni_anti-hierarchical_2004}. 
Figure \ref{fig:mdot} shows the resulting estimates of the average BH accretion rate within samples of galaxies as function of stellar mass and redshift, often referred to as an ``AGN main sequence'' \citep{mullaney_hidden_2012}. 
In general, our measurements of  $\langle \dot{M}(M_{\star},z) \rangle$ are in good agreement with previous measurements by \citet{yang_linking_2018}, increasing with $M_\star$ and with redshift (i.e. more BH growth occurring in galaxies of higher $M_\star$ and at higher $z$), although we note that we find somewhat higher absolute values of BH mass accretion in the most massive galaxies at our lowest redshifts.

We can now use these estimates of the average BH \emph{growth rates} to estimate the BH \emph{mass} that is grown through accretion over a given interval in redshift (i.e. cosmic time). 
The BH mass grown via accretion (for the typical galaxy of given $M_\star$ and $z$) is given by 
\begin{eqnarray}
    \Delta M_{BH}(M_{\star},z) &=& \langle\dot{M}(M_{\star},z)\rangle\times  \,\Delta t(z)
    \label{bhmass_gain}
\end{eqnarray}
where $\Delta t(z)$ corresponds to the interval in cosmic time corresponding to our chosen redshift ranges. We boost the accretion rate by a factor of 2 to account for additional accretion that is not directly observable in the X-ray, such as lower radiative efficiency accretion processes, and the up to $\sim$50\% of AGN that are heavily obscured (Compton-thick) and thus the observed X-ray luminosity would be substantially suppressed and such growth would be underestimated in our measurements \citep[e.g.][]{buchner_obscuration-dependent_2015,ananna_accretion_2019,carroll_high_2023}. 

We can also estimate the amount that the stellar mass of a galaxy will increase by over the same time interval, based on the average SFR. The SFRs of galaxies in our samples are based on the SED fitting described in Section \ref{sec:SED}. The average SFR is calculated for the full optically-selected parent galaxy populations in a given $M_{\star}-z$ bin, including any X-ray detected sources. The increase in $M_\star$ is thus given by
\begin{eqnarray}
    \Delta M_{\star}(M_{\star},z) &=& \langle SFR(M_{\star},z)\rangle\times(1-\mathcal{R})\times \,\Delta t(z)
    \label{stellarmass_gain}
\end{eqnarray}
where we assume a ``returned mass fraction'' $\mathcal{R}=40$\% and thus that only 60$\%$ of the average SFR is retained to build the galaxy stellar mass, while the remainder is returned instantaneously due to stellar winds and in supernovae after the death of short-lived massive stars \citep{madau_cosmic_2014}.
Since the average SFR and accretion rate depend on the stellar mass, we update them at each step of the tracking process to remain in the correct stellar mass bin if the new $M_{\star}$ has increased sufficiently to move to a higher mass bin after adding the mass gained.

In Figure \ref{fig:cml_mbh}, we show how much an average BH would grow between redshift $z=4$ and $z=0.1$ depending on the mass of its host galaxy, computing the mass gained using Equation \ref{bhmass_gain}. 
Leftmost points are the BH mass accumulated between $z=4-3$, and we then add on top of this initial mass, the accreted BH mass during every time step, keeping the same redshift binning as per Figure \ref{pledd}. Parallel to this process, we also grow the three average galaxies in stellar mass from initial masses of $\log M_{\star,start} = 10$, 10.5 and 11 $\rm M_\odot$.
We use their final stellar mass reached at the end of the growing process ($M_{\star,end}$) to estimate the expected BH mass at $z=0$ from the late-type scaling relation in \cite{greene_intermediate-mass_2020}. This expected BH mass
represents how massive a local BH should be in a galaxy of mass $M_{\star,end}$ at $z=0$. The difference between the rightmost point of the curves in Figure~\ref{fig:cml_mbh} and these BH mass estimates indicate that we are not able to account for the all of the BH mass at $z=0$ based on the \emph{accreted} BH mass between $z=4-0.1$. The numbers given at the side of Figure~\ref{fig:cml_mbh} specify the fraction of the expected BH mass at $z=0$ that is accounted for based on our estimates of the accretion grown since $z=4$. For massive galaxies $\geqslant 10^{10}\, \rm M_{\odot}$, only $\approx$ 19 - 33$\%$ of the final BH mass has been accreted since $z=4$, meaning that close to the entirety of their mass must have already been assembled prior to $z=4$ to produce those very massive BHs of mass $M_{BH} \sim 10^{8.5}\, \rm M_{\odot}$.

We also explore the tracks followed by BHs and galaxies in the $M_{BH}$--$M_\star$ space due to their BH accretion and stellar mass growth. We derive BH-galaxy growth tracks using four methods, shown in Figure \ref{fig:tracks}. In \textbf{method A}, we start galaxies at the beginning of the tracking process with $\log M_{\star,start} = 10, 10.5, 11$ at a range of different redshifts and then use the corresponding average SFRs to calculate the total growth in stellar mass over each redshift interval until we reach $z\approx0$. We assume these galaxies all start with a small BH mass ($M_{BH,start} = 10^{5-5.5-6} M_{\odot}$, depending on $M_{\star,start}$ although the exact choice has a negligible impact) and then add the relevant BH mass growth via accretion over the same redshift intervals. This results in the growth pathways shown in the top panels of Figure \ref{fig:tracks}, i.e, tracking forward from a small BH in a similar manner to the estimates in Figure~\ref{fig:cml_mbh}. We calculate six versions of these tracks, by starting the tracking process ($M_{\star,start} ; M_{BH,start}$) at different redshifts ($z_{start} = 4,3,2.5,2,1.5,1$). The rightmost point of each track is the final BH-galaxy mass at $z=0.1$. We also show the $M_{BH}-M_{\star}$ scaling relations from \cite{greene_intermediate-mass_2020} (late-type in brown and early-type in orange), as a reference for the final BH-galaxy mass, to show the measured relation (and corresponding intrinsic scatter) for observed, inactive BHs at $z=0$.

In \textbf{method A}, we see that the BH-galaxy evolution is very different depending on the choice of $M_{\star,start}$. With increasing $M_{\star,start}$, galaxy growth slows down while BHs are growing faster than in lower mass hosts, shown by the tracks getting more vertical. Indeed, in lower mass galaxies of $M_{\star,start} = 10^{10}\, \rm M_{\odot}$, we observe rapid growth in both the host and the BH, especially at higher redshift (red segment), before letting the BH drive those growth tracks alone at later times. However, for higher starting stellar mass of $M_{\star,start} = 10^{11}\, \rm M_{\odot}$, we observe strong vertical growth tracks across all redshifts, meaning that only the central BH is growing in a galaxy that has almost already reached its final mass. Such tracks show that significant BH growth may still occur at late cosmic times in very massive, primarily quiescent galaxies \citep[see also][]{aird_agn_2022,ni_incidence_2023}. 
Comparing the end point of these forward tracks with the scaling relations, we see that we can account for the entire BH mass expected in some massive late-type galaxies, as most, if not all, tracks fall within the observed intrinsic scatter of the late-type $M_{BH}-M_{\star}$ scaling relation. However, through the total integrated mass gained since $z=4$, we cannot account for enough mass to reach the median late-type scaling relation and thus cannot explain the majority of the BH masses of local, massive late-type galaxies. Moreover, taking into account the large dispersion of both scaling relations, the missing BH mass is even larger for early-type galaxies at fixed stellar mass, needing to reach at least one order of magnitude higher in BH mass.

In \textbf{method B}, we now start galaxies at the same $\log M_{\star,start} = 10, 10.5, 11$ at a range of different redshifts but this time seed them with a BH of mass derived through the scaling relations such that $M_{BH,start} = f(M_{\star,start})$, with $f$ being the scaling relations derived in \cite{greene_intermediate-mass_2020}. From these starting points, we grow the BH and galaxy forwards in time, following the same process as in method A. 
On the late-type relation, i.e. star-forming, (brown shaded region), we see that both BH and galaxy grow at comparable rates, with the galaxy growing faster at higher redshift, which is then taken over by the BH at later cosmic times. This sort of back and forth in the growth history of BH-galaxy, produces evolutionary tracks that roughly follow the median scaling relation and thus account at each time step, for most of the necessary BH growth. However, when looking at the quiescent population (orange region), hosting a higher mass BH at fixed stellar mass than the star-forming galaxies, we no longer observe significant BH growth relative to the already-high starting BH mass. We estimate a moderate amount of galaxy growth, producing a mostly flat curve. To avoid over-growing our galaxies on the quiescent branch, we computed the SFR from the main sequence relation in \cite{popesso_main_2022}, and removed 1 dex in order to suppress the average SFR we computed from the whole galaxy population in our sample, without distinguishing quiescent from star-forming galaxies. However, even when accounting for a smaller stellar mass increase in quenched galaxies, we still find that our galaxies drift away from the scaling relation, producing under-massive BHs at $z=0$.

Methods A and B showed that we are missing a significant fraction of the final expected BH mass when tracing BH growth through accretion across redshift $z=0.1-4$, particularly when considering the higher mass BHs in early-type galaxies. In methods C and D, we now account for this gap, starting the tracking process with an already fully grown massive BH at $z=0$ and working backwards in time to see how much BH mass we can account for over our range of redshift and how much BH mass must \emph{already} be in place at the beginning of the tracking process.

In \textbf{method C}, we adopt the final stellar masses, $M_{\star,end}$, calculated with Method A, but assume that the final ($z=0$) BH mass lies on either the early-type or late-type scaling relation. From these points, we now trace BH-galaxy assembly \emph{backwards} in time, removing BH or stellar mass at each time step, computed as per Equations \ref{bhmass_gain} and \ref{stellarmass_gain}. The same process is employed in \textbf{method D}, but this time assuming $M_{\star,start}$ at $z=0$ and $M_{BH,start} = f(M_{\star,start})$, as in method B but working backwards in time. On the quiescent branch, we use the same process as in method B to suppress the average SFR using the relation derived by \cite{popesso_main_2022} and removing 1 dex.  With these methods, we produce very different BH growth tracks. Our estimates of the mass assembled through accretion correspond to a relatively small fraction of the assumed $z=0$ $M_{BH}$, resulting in mainly flat tracks in this logarithmic space. Indeed, the backward tracks flatten more and more as we consider higher final BH masses. Looking at the upper-most tracks ($M_{BH} \sim 10^9\, \rm M_{\odot}$), we notice that the BH mass at high redshift (leftmost point), is roughly the same as the final BH mass ending on the scaling relation (rightmost point), meaning that only a small proportion of the final BH mass was grown during the $z=4$ to $z=0$ time period.
These objects must have assembled close to the entirety of their mass, prior to redshift $z=4$.

By measuring instantaneous SFR, BH growth rates, and current BH masses for a large sample of broad-line AGN at $z<0.35$, \cite{zhuang_evolutionary_2023} found similar growth patterns with fast-growing BHs falling below the scaling relation and little to no mass gain in the massive BH population above \citep[see also][]{sun_evolution_2015}.
Our results give a more comprehensive picture using direct measurements of the AGN fraction in massive galaxies and accounting for the average amount of BH growth over time, given that AGN are not a sustained phase over long timescales. 
Recent work by Terrazas et al. (in prep), adding BH growth based on observed sBHAR distributions to an empirical galaxy evolution model \citep{behroozi_universemachine_2019}, also identifies these differing pathways of BH growth.

\begin{figure}
    \centering
    \includegraphics[width=\columnwidth]{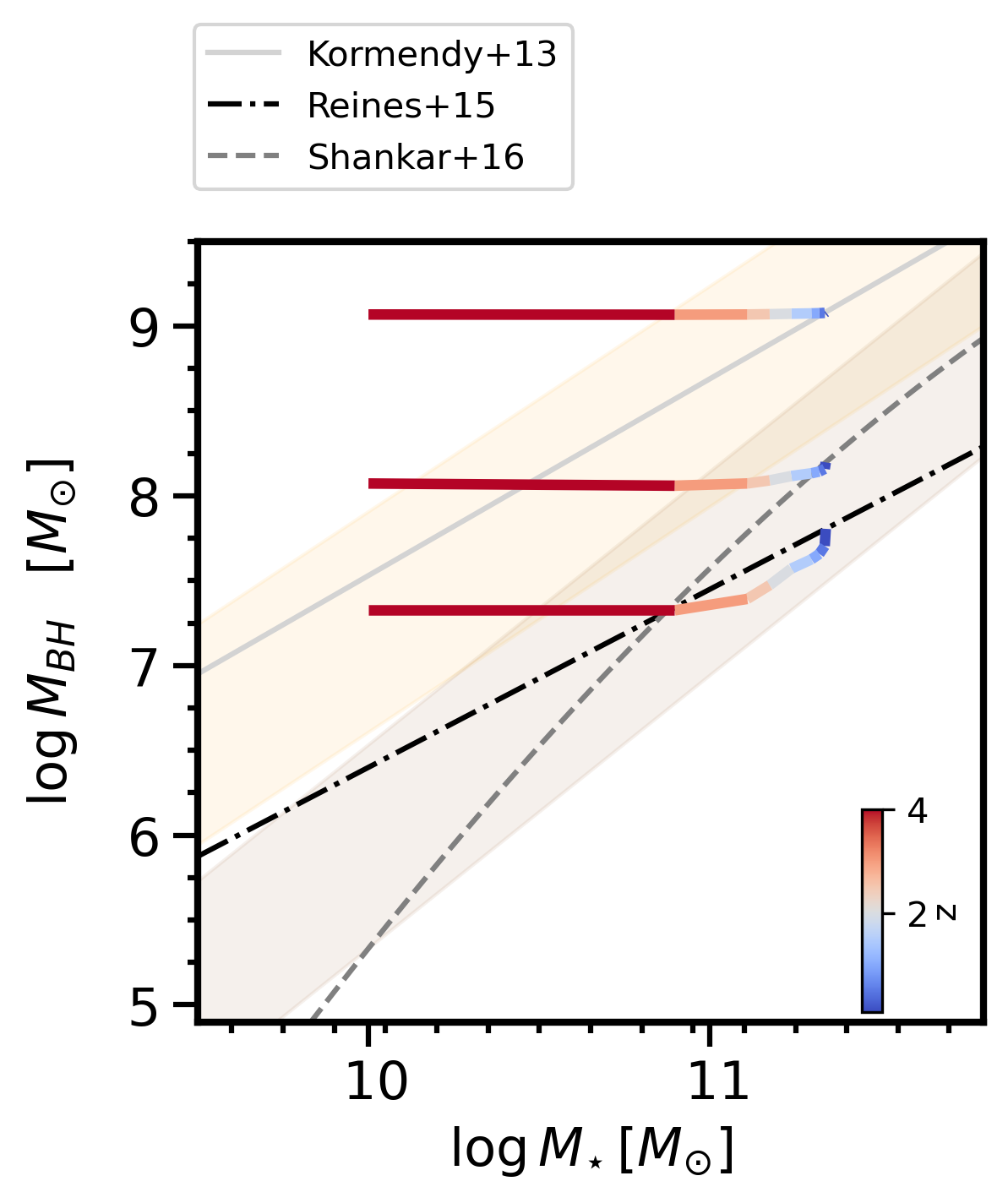}
    \caption{Black Hole - galaxy growth tracks using Method C for a starting galaxy of $M_{\star , start} = 10^{10}\, \rm M_{\odot}$ at $z=4$ (as in the leftmost panel of the third row of Figure \ref{fig:tracks}) adopting different $z=0$ scaling relations. We show different final BH mass at $z=0$ as expected by \citet{kormendy_coevolution_2013} as the solid grey, \citet{reines_relations_2015} as the black dot-dashed line and \citet{shankar_selection_2016} as the dashed grey line. The orange and brown shaded areas show the dispersion in the early- and late-type galaxy relations of \citet{greene_intermediate-mass_2020}, as in Figure~\ref{fig:tracks}.}
    \label{fig:scaling_comp}
\end{figure}

The small fraction of missing BH mass at $z=0$ for the lower BH mass population and lack of sufficient growth in the high mass population could be explained by different scenarios. First, we assume here that the expected final $M_{BH}$ follows strictly the median relation of \cite{greene_intermediate-mass_2020}, which is on average a reasonable assumption but could be biased towards higher mass BHs where we can accurately measure their mass. To explore this further, in Figure \ref{fig:scaling_comp} we compare a few of the main $M_{BH} - M_\star$ scaling relations from the literature i.e \cite{kormendy_coevolution_2013, reines_relations_2015, shankar_selection_2016}. Here, we re-compute the BH-galaxy growth tracks using Method C in a galaxy of $M_{\star ,start} = 10^{10}\, \rm M_\odot$ at $z=4$, i.e 3rd line - 1st column panel of Figure \ref{fig:tracks}, to trace back in time the growth of a local BH in a galaxy of mass $M_{\star ,start}$ according to each of the scaling relations. We see that in the high stellar mass regime ($M_\star \geqslant \, 10^{11}\, \rm M_\odot$) at $z=0$, both the \cite{kormendy_coevolution_2013} and \cite{shankar_selection_2016} relation yield similar results to the \cite{greene_intermediate-mass_2020} early-type and late-type relations respectively, confirming the need for an early assembly of the bulk of BH mass. There are  larger discrepancies in the lower stellar mass regime between the \cite{shankar_selection_2016} relation (grey dashed line) and the late-type scaling relation from \cite{greene_intermediate-mass_2020} (brown region), which does not affect the massive BH population targeted in this paper. Finally, when tracking BH-galaxy assembly starting from the \cite{reines_relations_2015} relation, we can recover a slightly larger fraction of the total BH mass between $z=0-4$, which is reflected in the larger relative BH mass increase at later times. Indeed, they find a shallower relation and thus the typical BH mass is relatively lower in more massive galaxies. However, this scaling relation is partly based on an AGN sample and may be picking up galaxies currently undergoing relatively rapid growth in their BHs and thus moving up from below the median scaling relation. Overall, most scaling relations agree that on average, the assembly of the most massive BHs require a rapid and early BH growth, and although lower median scaling relations exist, all studies find a large intrinsic scatter and they do not refute the existence of the highest mass BHs detected at $M_{BH} \geqslant \, 10^8 \, \rm M_\odot$ that do require an almost complete assembly prior to $z=4$.

Moreover, we did not account for BH growth through mergers as it is believed to have a drastically lower contribution to BH assembly compared to accretion, but could still play a part in this missing BH mass and even become the dominant growth channel for BHs, especially at lower redshift ($z<2$) in massive populations \citep{pacucci_separating_2020}. We also assumed a simple galaxy growth model through star formation with a returned fraction of 40$\%$, however mergers could also play a significant part in the late assembly of galaxies, especially for massive galaxies ($M_\star \gtrsim 10^{11}\, \rm M_{\odot}$) where a large fraction of the stars may have formed \textit{ex-situ} and contribute to the galaxy stellar mass through mergers \citep{husko_buildup_2022}. Refining the growth history and accounting for mergers in both BH and galaxy assembly could be a route to provide this missing growth between local massive BHs and our average BH population, but is deferred to future work.

Our study suggests that massive BHs assembled early, possibly even before the assembly of galaxy stellar mass; that much of this growth happened at $z>4$, consistent with recent JWST results revealing large populations of high-redshift AGN and indicating high levels of early BH growth \citep[e.g.]{maiolino_jades_2023,ubler_ga-nifs_2023,marshall_ga-nifs_2023, harikane_jwstnirspec_2023}; and that BHs have assembled through a sustained extreme accretion, approaching or exceeding the Eddington limit.

\section{Conclusions}
\label{sec:conclusion}

In this paper we have presented different measurements of the X-ray AGN incidence in massive galaxies using the \textit{Chandra} Deep-Wide Field Survey (CDWFS) in the Boötes field. Our main conclusions are as follows:

\begin{itemize}

    \item We find essentially no correlation between stellar mass, $M_{\star}$, and the instantaneous X-ray luminosity, $L_X$. This shows that a galaxy of given mass can produce widely different X-ray luminosities, confirming that the processes responsible for the X-ray AGN emission must vary on much shorter timescales than galaxy growth mechanisms. However, we do find a weak trend whereby the most X-ray luminous AGN are only found in the highest stellar-mass host galaxies ($M_\star\gtrsim10^{10.5}\, \rm M_\odot$). 

    \item 
    Measurements of the fraction of galaxies hosting AGN above fixed limits in X-ray luminosity show a strong increase as a function of stellar mass. 
    This dependency is mostly caused by a well-understood observational bias whereby BHs accreting at similar rates (relative to the host galaxy stellar mass) produce a higher observable luminosity.
    We account for this observational bias by instead measuring AGN fractions to fixed limits in specific black hole accretion rate (sBHAR, $L_X/M_\star$). After accounting for this bias, we do not find evidence for a strong stellar mass dependence in the high mass regime, indicating that galaxies of all masses above $M_{\star} \geqslant 10^{10}\, \rm M_{\odot}$ at a given redshift are, to first order, equally likely to host AGN.  

    \item We find a strong increase in the AGN fraction (to sBHAR limits) with increasing redshift as well as an increase in the average sBHAR $\langle\lambda\rangle$ spanning values from $\langle\lambda\rangle \sim 10^{31.2 - 32.8}\, \rm erg/s/M_{\odot}$ between $z = 0.1-4$ and thus reaching almost $10\%$ of the Eddington limit on average in the early Universe. Additionally, the AGN fraction at low sBHAR limits appears slightly enhanced in the most massive galaxies while the pattern reverses for $\lambda > 10^{34}\, \rm erg/s/M_{\odot}$. This indicates that substantial BH growth took place at earlier times, and that extreme accretion events may be suppressed within the most massive galaxies.

    \item Based on our measurements of BH accretion rates, we track the growth of both BH mass and galaxy stellar mass over cosmic time for different galaxy populations. 
    We are able to account for the bulk of the mass assembly needed to produce the relatively low-mass BHs found at $z=0$ ($M_{BH} \sim 10^{6 - 7.5}\, \rm M_{\odot}$, corresponding to BH masses below the median $M_{BH}-M_{\star}$ scaling relation for late-type /star-forming galaxies) based on the observed accretion growth between $z=4$ and $z=0.1$ in the likely progenitors of such galaxies. 
   
    \item We find that to produce the most massive BHs in the local Universe ( $M_{BH} \gtrsim \, 10^8 \, \rm M_{\odot}$) that most of their mass must have already been assembled prior to $z=4$. Once reaching these extreme masses, our measurements suggest little substantial additional mass growth (in relative terms) occurs between $z=4$ and present times.
    
\end{itemize}

Through the study of 2114 hard X-ray selected AGN over a large optical sample of galaxies in the Boötes field, we quantified the amount of BH assembly across cosmic time and how it proceeds alongside their galaxies. Our results show that while low-mass local BHs can mostly be grown through accretion from $z=4$, this growth mechanism struggles to produce the most massive BHs observed at $z=0$. This indicates the importance and necessity of accurately measuring the amount of BH growth occurring in different galaxy populations over cosmic time and in particular to study the putative rapid build-up of massive black holes at $z>4$, as well as quantifying the degree of contribution of BH assembly through mergers.

\section*{Acknowledgements}
PG acknowledges funding from an STFC studentship. JA and ALR acknowledge support from a UKRI Future Leaders Fellowship (grant code: MR/T020989/1). AG acknowledges funding from the Hellenic Foundation for Research and Innovation (HFRI) project ‘4MOVE-U’ ("Forward-Modeling the Variable Energetic Universe") grant agreement 2688, which is part of the programme 2nd Call for HFRI Research Projects to support Faculty Members
and Researchers. The scientific results reported in this article are based to a significant degree on observations made by the \textit{Chandra} X-ray Observatory. For the purpose of open access, the authors have applied a Creative Commons Attribution (CC BY) licence to any Author Accepted Manuscript version arising from this submission.
\section*{Data Availability}
Optical data used in this work come from the multi-wavelength catalogue presented by \citet{kondapally_lofar_2021} and the X-ray catalogue can be found in \cite{masini_chandra_2020}. The derived measurements are available from the corresponding author by request.



\bibliographystyle{mnras}
\bibliography{xbootes} 




\appendix

\section{SED fitting parameters}

In this Appendix, we share the full parameter grid used in the SED fitting with \textsc{cigale} for both models, the optically and X-ray selected populations (see Table \ref{cigale_param}).

\begin{table*}
	\centering
	\caption{\textsc{cigale} parameter values used in each module. This grid leads to 1,134,000 and 251,942,400 models for the optical and X-ray samples respectively.}
	\label{cigale_param}
	\begin{tabular}{lcr} 
		\hline
		Module & Parameter & Value\\
		\hline
		\multirow{6}{16em}{\ascii{sfhdelayed}} & \ascii{tau\_main} ($\rm Myr$) & 1,50,200,800,1000\\
                           & \ascii{age\_main} ($\rm Myr$) & 100, 200,1000, 2000, 3000, 4000\\
                           & \ascii{tau\_burst} ($\rm Myr$) & 50\\
                           & \ascii{age\_burst} ($\rm Myr$) & 20\\
                           & \ascii{f\_burst} & 0\\
                           & \ascii{sfr\_A} & 1\\
        \hline
		\multirow{3}{16em}{\ascii{bc03}} & \ascii{imf} & 1 (Chabrier)\\
                                        & \ascii{metallicity} & 0.02\\
                                        & \ascii{separation\_age} ($\rm Myr$) & 10\\
        \hline
		\multirow{6}{16em}{\ascii{nebular}} & \ascii{logU} & -2\\
                                            & \ascii{zgas} & 0.02\\
                                            & \ascii{ne} & 100\\
                                            & \ascii{f\_esc} & 0.0, 0.02, 0.5\\
                                            & \ascii{f\_dust} & 0\\
                                            & \ascii{lines\_width} ($\rm km.s^{-1}$) & 300\\
		\hline
		\multirow{5}{16em}{\ascii{dustatt\_modified\_CF00}} & \ascii{Av\_ISM} & 0,0.1,0.2,0.5,1.0,1.5,2,2.5,3\\
                                                            & \ascii{mu} & 0.44\\
                                                            & \ascii{slope\_ISM} & -0.7\\
                                                            & \ascii{slope\_BC} & -1.3\\
                                                            & \ascii{filters} & V\_B90 \& FUV\\
		\hline
  		\multirow{4}{16em}{\ascii{dl2014}} & \ascii{qpah} & 2.5\\
                                            & \ascii{umin} & 1\\
                                            & \ascii{alpha} & 2\\
                                            & \ascii{gamma} & 0.1, 0.9\\
		\hline
        \multicolumn{3}{|c|}{X-ray sample added components} \\
        \hline
  		\multirow{15}{16em}{\ascii{skirtor2016}} & \ascii{t} & 5, 7\\
                                                 & \ascii{pl} & 1\\
                                                 & \ascii{q} & 1\\
                                                 & \ascii{oa} & 10, 40, 80\\
                                                 & \ascii{R} & 20\\
                                                 & \ascii{Mcl} & 0.97\\
                                                 & \ascii{i} & 30, 70, 90\\
                                                 & \ascii{disk\_type} & 0 (Skirtor)\\
                                                 & \ascii{delta} & -0.36\\
                                                 & \ascii{fracAGN} & 0,0.1,0.2,0.5,0.8,0.99\\
                                                 & \ascii{lambda\_fracAGN} & 0/0\\
                                                 & \ascii{law} & 0 (SMC)\\
                                                 & \ascii{EBV} & 0,0.1,0.2,0.3\\
                                                 & \ascii{temperature} (K) & 100\\
                                                 & \ascii{emissivity} & 1.6\\
		\hline
  		\multirow{7}{16em}{\ascii{xray}} & \ascii{gam} & 1.9\\
                                        & \ascii{E\_cut} & 300\\
                                        & \ascii{alpha\_ox} & -1.9, -1.7, -1.5, -1.3, -1.1\\
                                        & \ascii{max\_dev\_alpha\_ox} & 0.2\\
                                        & \ascii{angle\_coef} & 0.5 \& 0\\
                                        & \ascii{det\_lmxb} & 0\\
                                        & \ascii{det\_hmxb} & 0\\
		\hline
	\end{tabular}
\end{table*}


\bsp	
\label{lastpage}
\end{document}